\documentclass[12pt,draftclsnofoot,onecolumn]{IEEEtran}
\normalsize

\usepackage{graphicx}
\usepackage{epstopdf}

 \DeclareGraphicsExtensions{.pdf}

\usepackage[cmex10]{amsmath}
\usepackage{bm}
\usepackage{epsfig}
\usepackage{booktabs} 
\usepackage{latexsym}
\usepackage{enumerate}
\usepackage{float} 
\usepackage{subfig}
\usepackage{multirow}
\usepackage{epstopdf}
\usepackage{longtable}
\usepackage{stfloats}
\usepackage{color} 
\usepackage{amssymb}
\graphicspath{{./Figures/}}
\usepackage{color}
\usepackage{tikz}
\usepackage{bbm}
\usepackage{authblk}
\usepackage{verbatim}
\usepackage{mathtools}

\usepackage{balance}

\usepackage{siunitx}
\DeclareSIUnit\bps{bps}
\DeclareSIUnit\Torr{Torr}
\DeclareSIUnit\torr{Torr}
\DeclareSIUnit\sample{Sa}
\DeclareSubrefFormat{parens}{#1(#2)}
\usepackage{cite}

\usepackage{color, soul}
\newcommand{\rev}[1]{\textcolor{black}{{#1}}}

\begin{document}
% TO COUNT LINES PER PAGE
%\setpagewiselinenumbers
%\pagewiselinenumbers 

\title{Channel Measurement and Ray-Tracing-Statistical Hybrid Modeling for Low-Terahertz Indoor Communications}

\author{Yi~Chen,~\IEEEmembership{Student~Member,~IEEE}, Yuanbo~Li, Chong~Han,~\IEEEmembership{Member,~IEEE}, Ziming~Yu, and Guangjian~Wang 
\vspace{0cm}

\thanks{Y. Chen, Y. Li and C. Han are with the Terahertz Wireless Communications Laboratory, University of Michigan-Shanghai Jiao Tong University Joint Institute, Shanghai Jiao Tong University, Shanghai, China (email: \{yidoucy, yuanbo.li, chong.han\}@sjtu.edu.cn).}
\thanks{Z. Yu and G. Wang are with Huawei Technologies Co., Ltd, China (email: \{yuziming, wangguangjian\}@huawei.com).}
%\thanks{Corresponding author:  Chong Han, e-mail: chong.han@sjtu.edu.cn}
}

\maketitle

\begin{abstract}
TeraHertz (THz) communications are envisioned as a promising technology, owing to its unprecedented multi-GHz bandwidth. One fundamental challenge when moving to new spectrum is to understand the science of radio propagation and develop an accurate channel model. In this paper, a wideband channel measurement campaign between 130 GHz and 143 GHz is investigated in a typical meeting room. Directional antennas are utilized and rotated for resolving the multi-path components (MPCs) in the angular domain. With careful system calibration that eliminates system errors and antenna effects, a realistic power delay profile is developed. Furthermore, a combined MPC clustering and matching procedure with ray-tracing techniques is proposed to investigate the cluster behavior and wave propagation of THz signals. In light of the measurement results, physical parameters and insights in the THz indoor channel are comprehensively analyzed, including the line-of-sight path loss, power distributions, temporal and spatial features, and correlations among THz multi-path characteristics. Finally, a hybrid channel model that combines ray-tracing and statistical methods is developed for THz indoor communications. Numerical results demonstrate that the proposed hybrid channel model shows good agreement with the measurement and outperforms the conventional statistical and geometric-based stochastic channel model in terms of the temporal-spatial characteristics.
%TeraHertz (THz) communications are envisioned as a key technology for the next wireless systems, owing to its unprecedented multi-GHz bandwidth. 
%One fundamental challenge when moving to new spectrum is to understand the science of radio propagation and develop an accurate channel model.
%In this paper, a wideband channel measurement campaign between 130~GHz and 143~GHz is investigated in a typical meeting room for promising THz wireless communication networks. Directional antennas are utilized and rotated for resolving the multi-path components (MPC) in the angular domain. With careful system calibration that eliminates system errors and antenna effects, the realistic power delay profile is developed. Furthermore, a combined MPC clustering and matching procedure with ray tracing techniques is proposed to further investigate the cluster behavior and the wave propagation of THz signals. In light of the measurement results, physical parameters and insights in the THz indoor channel are comprehensively analyzed, including the line-of-sight path loss, power distributions, temporal and spatial features, and correlations among THz multi-path characteristics. Finally, a hybrid channel model that combines ray tracing and statistical channel modeling is developed for THz indoor communications. Numerical results suggest that the proposed hybrid channel model shows good agreement with the measurement and outperforms the conventional statistical channel model and geometric-based stochastic channel model in terms of the temporal-spatial channel characteristics.
\boldmath

\end{abstract}
\begin{IEEEkeywords}
Terahertz communications, Channel measurement, Temporal-Spatial analysis, Ray-Tracing-Statistic hybrid modeling.
\end{IEEEkeywords}
\section{Introduction}
In the past few decades, wireless data traffic has witnessed exponential growth and brings an increasing demand for high-speed wireless communication. New spectral bands are required to support Terabit-per-second (Tbps) data rates for future wireless applications~\cite{akyildiz2014terahertz}. Currently, wireless local area networks (WLAN) techniques, i.e., 802.11ad protocol and fifth-generation (5G) mobile networks, have opened up the millimeter-wave (mmWave) spectrum (10-100~GHz) to seek for broader bandwidth and higher data rates~\cite{shafi2018microwave,wang2018survey}. However, still limited to several GHz bandwidth, the mmWave cannot support Tbps requirements. To further move up the carrier frequency, the Terahertz (THz) band spanning over 0.1 and 10~THz, is envisioned as one of the promising spectrum bands to enable ultra-broadband 6G communications. With tens and hundreds of GHz contiguous spectrum resources, the use of THz band can address the spectrum scarcity and capacity limitations of current wireless systems.
To make THz communications come into reality in the very near future, the first spectral window of the THz band is centered at 140~GHz band, which can be implemented by mature electronic-based device technologies~\cite{xing2018propagation}.

When moving to the THz spectrum, fundamental challenges to designing THz wireless systems include understanding the science of radio propagation, development of efficient yet accurate channel models, and thorough analysis of channel characteristics in the THz band.
To understand the science of radio propagation, and develop channel models to describe major propagation processes, three main approaches are adopted, including physical measurement using a channel sounder~\cite{guan2020channel}, geometric-optic method~\cite{chong2017thz}, and full-wave EM field analysis~\cite{chen2019channel}. In this paper, we adopt physical channel measurement to acquire the realistic THz propagation features between the transmitter and receiver in a meeting room.
\par \rev{From the literature, a number of channel measurement campaigns at THz frequencies have been reported for short-distance indoor scenarios~\cite{priebe2011channel,kim2015d,kim2016characterization,eckhardt2019measurements,xing2019indoor,fu2020modeling,cheng2020thz}, outdoor vehicular communications~\cite{guan2016millimeter,abbasi2019double,petrov2020measurements}, and  aircraft cabin communications~\cite{eckhardt2020indoor}. Most of these studies are above 300 GHz, which is higher than 140 GHz that we concern. The channel measurement at 140 GHz for indoor scenarios includes short-range on a desk~\cite{kim2015d}, and in-building~\cite{xing2019indoor}. In particular, these studies mainly focus on the analysis of path loss, reflection loss, and penetration loss~\cite{kim2015d,xing2019indoor}.
However, an extensive channel measurement with various transmitter (Tx) and receiver (Rx) positions in a meeting room environment for 140~GHz is still missing.} Furthermore, thorough channel characterization on the temporal-spatial THz propagation features as well as large-scale fading path loss is not well captured, which includes K-factor, sparsity, temporal and spatial dispersion, correlation of multi-path features, among others.

\par In this paper, we present a wideband channel measurement campaign based on the vector network analyzer (VNA) at 130-143~GHz in a typical meeting room. With two fixed Tx, multi-path measurements are conducted at ten different receiver positions. Directional antennas are equipped at Tx and Rx to resolve multi-path components (MPCs) in the angular domain. \rev{To assist post-processing of the measured data, we invoke ray-tracing (RT) techniques to simulate the MPCs with detailed propagation trace in the same geometry where our channel measurement is implemented. Then, a combined MPCs clustering and matching procedure with a multi-path component distance (MCD) based Density-Based Spatial Clustering of Applications with Noise (DBSCAN) is proposed to cluster the measured MPCs and label the MPC clusters based on the RT results. }

In light of the measurement results, we characterize the physical parameters and provide insights of the THz indoor channel. 
Specifically, the measured path loss of the line-of-sight (LoS) path is validated with theoretical computation results.
Moreover, the power-delay-angular profiles (PDAPs) are presented to reveal the multi-path channel sparsity in the THz indoor environment and the significance of reflection over walls.
Furthermore, temporal and spatial features, including the K-factor, delay, and power spreads, are analyzed.
A correlation matrix is formulated to analyze the dependence among the multi-path characteristics of the THz indoor channel, e.g., Tx-Rx separation distance, number of clusters, K-factor, delay spread, and angular spread. At last, we develop a hybrid semi-deterministic channel model that combines the geometric-optic method (ray tracing) and statistical modeling method, which suggests that the non-line-of-sight (NLoS) path reflects over the walls are dominating in the channel. Numerical results show that the proposed hybrid model outperforms the conventional statistical and the 3rd Generation Partnership Project  (3GPP) geometric-based stochastic channel models (GSCM) models with the metric of PDAP.

Compared to our preliminary and shorter version~\cite{yu2020wideband}, this full-fledged work includes multi-path clustering effects and development of a RT-statistical hybrid channel model for the indoor THz channel, with substantially more performance analysis.
%For simulating the WLAN communication scenarios where an access point (AP) conducts beamforming and the main beam is aligned to the user equipment (UE), the transmitter is fixed at a corner of the room with a height of 2~m, and the transmitters are placed at ten different positions distributed in the whole meeting room with a height of 1.4~m. The beamwidth of Tx is $30^\circ$ for wide coverage. To obtain a high spatial resolution, the beamwidth of Rx is $10^\circ$, which is three times smaller than Tx. Rx is amounted to the rotation unit and scan in both the azimuth and elevation domain with the angular step of $10^\circ$ and $5^\circ$, respectively, for recording the angular distributions of MPCs. Therefore, ten Tx-Rx pairs are measured in the measurement campaign. Based on the measurement results, we analyze the large-scale fading, i.e., path loss, and temporal and spatial characteristics of MPCs according to the positions of Rx. Furthermore, we study the penetration loss and reflection loss of different materials.
The distinctive contributions of this work are summarized as follows.
\begin{itemize}
	\item We establish a wideband THz channel measurement campaign in a meeting room, between 130 GHz and 143 GHz via VNA.
	%Tx position was fixed, while Rx is located in different positions. 
	Especially, Rx with a directional antenna scans in both azimuth ($[0^\circ$, $360^\circ]$) and elevation domains ($[-20^\circ$, $20^\circ]$) with the rotation step of $10^\circ$ for resolving the MPCs in the spatial domain. \rev{27 Tx-Rx positions are measured, including 21 for channel analysis and model parameter extraction and 6 for model validation.}
	\item {An MPCs clustering and matching procedure with an MCD-based DBSCAN algorithm is proposed to cluster the MPCs and match the measured MPCs with the simulated MPCs with the RT simulator.} The proposed procedure outperforms conventional K-Power-Means (KPM) algorithm in our channel measurement data. 
	%And RT simulation collaborating with the proposed clustering and matching procedure is conducted in the measurement environment for understanding the propagation in the environment.
	\item A comprehensive analysis of the channel characteristics is provided, including the path loss, reflection loss, cluster behavior, K-factor, delay spread, and angular spread. The correlation matrix of the channel parameters is evaluated to discuss the temporal-spatial properties.
	\item An RT-statistical hybrid channel model for the indoor THz channel at 140 GHz is developed on the basis of the analysis of the channel characteristics. The proposed hybrid model has low complexity without requiring detailed geometry information of the environment. Delay and angular spreads of the proposed model are validated and shows good agreement with the channel measurement. 
	\item A measure of figure structure similarity, Structural Similarity Index Measure (SSIM), is introduced to measure the accuracy of the channel model along with the root mean square error (RMSE). SSIM and RMSE between the generated PDAP by the proposed hybrid model and measured PDAP are evaluated, respectively. Compared to the conventional statistical channel model and the 3GPP GSCM~\cite{3gpp2018study}, the proposed hybrid model shows improved performance in characterizing PDAP of the THz multi-path channel.
\end{itemize}
\par The remainder of this paper is organized as follows. In Sec. II, we describe the channel measurement campaign, post-processing of measured data, and an MPCs clustering and matching procedure. Then, we present the channel measurement results and, more importantly, thoroughly analyze the channel characteristics in Sec. III. Furthermore, an RT-statistical hybrid channel model is proposed and validated in Sec. IV. Finally, the paper is concluded in Sec. V.
\begin{figure}[htbp]
\centering
\includegraphics[width=0.45\textwidth]{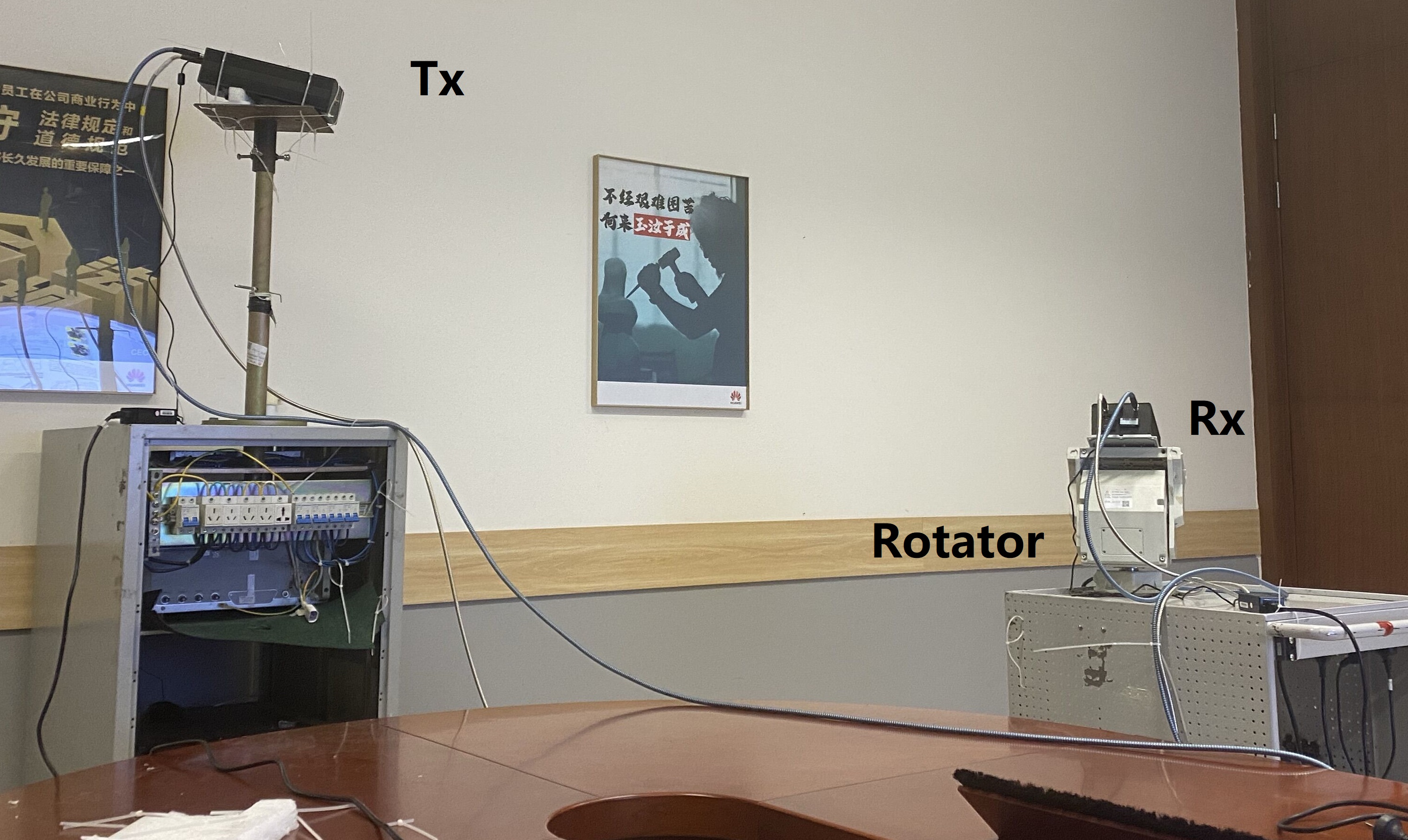}
\caption{Channel measurement platform.}
\label{fig:testpad}
\end{figure}
%%%%%%%%%%%%%%%%%%%%%%%%%%%%%%%%%%%%%%
\section{Channel Measurement Campaign}
%Before presenting the results of the channel measurement and the analysis of the THz indoor channel characteristics, we first detail the descriptions of the channel measurement campaign to help them better understood. 
	In this section, we describe the THz measurement campaign, including the specification of the hardware system, meeting room environment, and measurement deployment. Moreover, \rev{system calibration is carried out for eliminating the effects of the cables and radio-frequency (RF) fonts on the measured channel, as detailed in [1]. Ray-tracing techniques are used to trace the propagation of measured MPCs. We further propose a clustering and MPC matching procedure for analyzing MPCs clustering and match the measured data with the RT simulator output.}

\subsection{Channel Measurement System}
\par The measurement platform mainly consists of 140~GHz RF fronts with horn antennas and a VNA. The \rev{local oscillator (LO)} signal of 10.667 GHz is multiplied by a factor of 12 to 128 GHz, and the \rev{intermediate frequency (IF)} signals from 2 GHz to 15 GHz is up-converted by the multiplied LO signal to the frequency band from 130 to 143~GHz. As a result, the measured bandwidth $B_w$ is 13 GHz, and the time domain resolution, $\Delta \tau=1/B_w$, is 76.9~ps, which suggests we can resolve two paths if the distance difference between them is larger than 2.3~cm. The recorded points in the frequency domain or equivalently, the sweeping frequency points are 1301, which corresponds to the sweeping interval at $\Delta f=10$~MHz. The maximum excess delay, $\tau_m=1/\Delta f$, is calculated as 100 ns, and accordingly, the largest traveling distance of a detectable path to be measured is $L_m=30$~m, which is sufficient for channel measurement in a meeting room.
%\ch{CH: I remember reviewers asked about the numbers here. Are they provided by the measurement system?} \yi{I add some intermediate calculations. Those parameters are very common in the channel measurement paper and can be easily calculated.}

\par Both Tx and Rx are equipped with horn antennas, which are shown in Fig.~\ref{fig:testpad}. The antenna at Tx produces an antenna gain of 15 dBi with the half-power beamwidth (HPBW) of $30^\circ$ at 140~GHz, to enable a sufficient angular coverage. The Rx antenna gain is 25 dBi, and the HPBW is $10^\circ$, which is one-third of that at Tx for high spatial resolution. The two antennas are mounted on two rotation units, which can be rotated by step motors. In addition, the power of the test signal is 1 mW, and the noise floor of our THz measurement platform is $-160$ dBm. \rev{In our measurement, the maximum received signal strength is $-79.5$~dBm.} The detailed parameters of the measurement system are summarized in Table~\ref{tab:mparameters}.
%\ch{CH: can we have a figure (real hardware, or block diagram) of the measurement system?}\yi{I photographed one.}
% Table generated by Excel2LaTeX from sheet 'Sheet1.'
\begin{table}[htbp]
  \centering
  \caption{Parameters of the Measurement System.}
    \begin{tabular}{lll}
        \toprule
    Parameter & \multicolumn{1}{l}{Symbol} & Value \\
    \midrule
    Start frequency &   $f_{start}$    & 130~GHz \\
    End frequency &   $f_{end}$    & 143~GHz \\

    Bandwidth &   $B_w$   & 13~GHz \\
    Sweeping points &   $ N$  & 1301\\
    Sweeping interval &    $\Delta f$   & 10~MHz \\
    Average noise floor &  $P_N$     & -160~dBm \\
    Test signal power &   $P_{in}$    & 1~mW \\
    HPBW of transmitter & $HPBW^{Tx}$     & $30^\circ$ \\
    HPBW of receiver &  $HPBW^{Rx}$  & $10^\circ$ \\
    Antenna gain at Tx &   $G_{\text{t}}$   & 25 dBi \\
    Antenna gain at Rx &   $G_{\text{r}}$   & 15 dBi \\
    Time domain resolution &  $\Delta \tau$     & 76.9~ps \\
    Path length resolution &   $\Delta L$    & 2.3~cm \\
    Maximum excess delay &    $\tau_m$   & 100~ns \\
    Maximum path length &   $L_m$    & 30~m \\
    Azimuth rotation range &      & $[0^\circ:10^\circ:360^\circ]$ \\
    Elevation rotation range &    & $[-20^\circ:10^\circ:20^\circ]$ \\
        \bottomrule
    \end{tabular}%
  \label{tab:mparameters}%
\end{table}%

\subsection{Meeting Room Environment and Measurement Deployment}

\begin{figure*}

\centering
\subfloat[Meeting room for THz channel measurements.]{
\includegraphics[height=6cm,width=7.5cm]{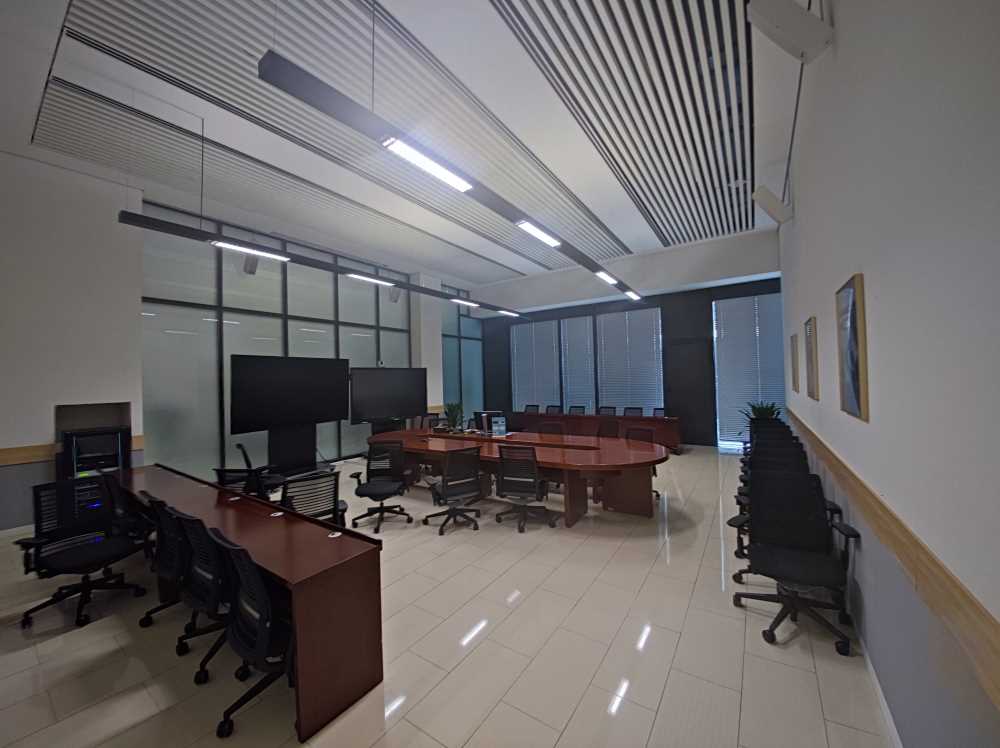}
\label{fig:meeting_room}
%\caption{fig1}
}
\centering
\subfloat[Overview of the meeting room and measurement deployment.]{
\includegraphics[height=6cm,width=8cm]{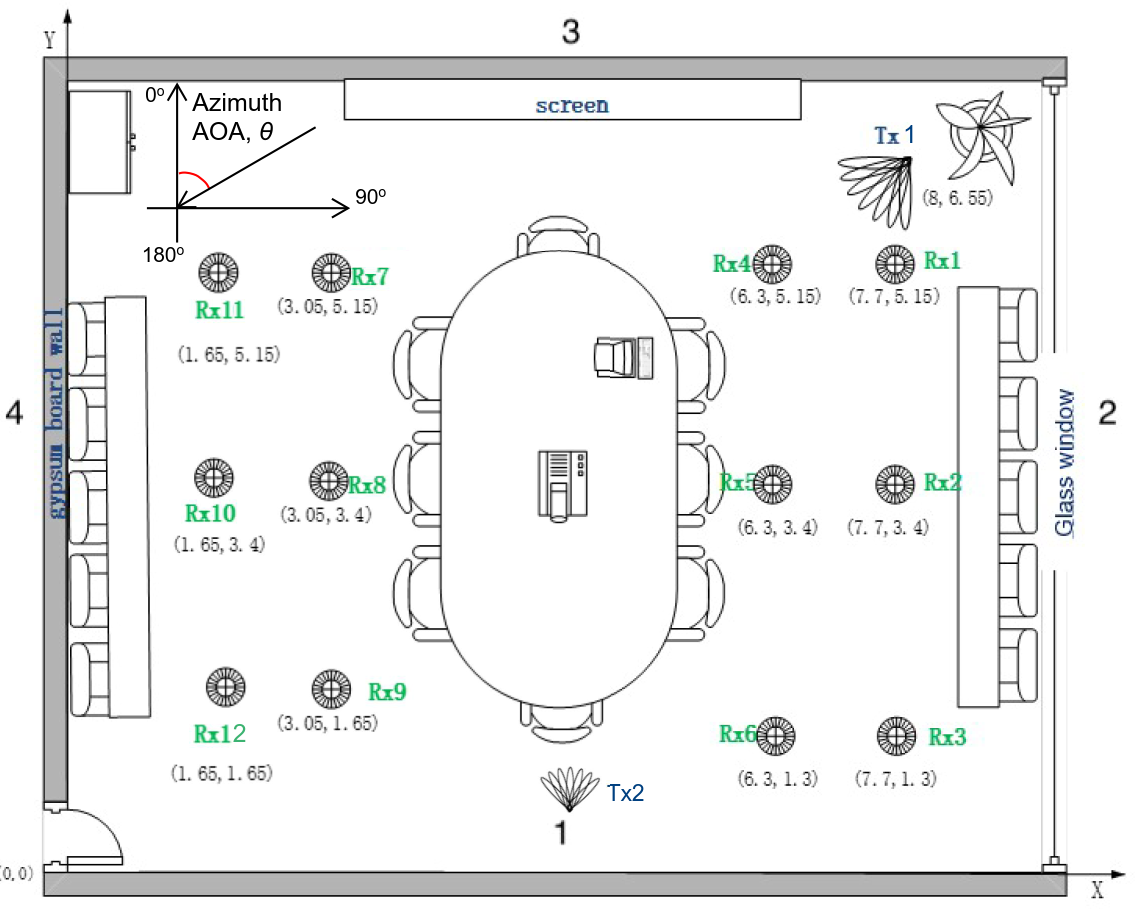}
\label{fig:deployment}
}
\caption{The meeting room layout and measurement deployment.}
\label{sec:deployment}
\end{figure*}
We carry out the channel measurement in a typical meeting room with an area of 10.15 m $\times$ 7.9 m and a ceiling height of 4 m. In the meeting room, a 4.8 m $\times$ 1.9 m desk with a height of 0.77 m is placed in the center, and eight chairs are around the desk, as shown in Fig.~\ref{fig:meeting_room}. In addition, two TVs are closely placed in front of a wall. The material of one wall is glass, while the other three are limestones. We notice that the maximum detectable path length imposed by the measurement system is 30~m, which is three times the dimension of the meeting room. As a result, reflected paths with at most three-order reflection can be recorded in our measurement. 
\par To simulate the indoor communication scenario, Tx with HPBW of $30^\circ$ is fixed as an AP at a corner of the meeting room. The Tx height is 2~m, which is higher than the Rx or UE with a height of 1.4~m. The wide HPBW for Tx guarantees the coverage of target UEs, while the small HPBW for Rx of $10^\circ$ provides high spatial resolution for spatial-domain channel measurement. In our measurement deployment, 2 positions of Tx as well as 12 positions of Rx are set in the meeting room,  as depicted in the top view of the meeting room in Fig.~\ref{fig:deployment}. Tx1 is at the corner of the meeting room while Tx2 is behind the table and near wall 1. In this measurement campaign, \textit{two measurement sets} are conducted. In the measurement set 1, Tx is placed on the position Tx1, the Rx is placed on the positions Rx1-4 and Rx6-10. In the measurement set 2, Tx is placed on the position Tx2 while Rx is placed on the positions Rx1-12. For the measurement of each Rx, the main beam of Tx is directed to the Rx. By contrast, Rx with the spatial resolution of $10^\circ$ scans the receiving beam in the azimuth domain from $0^\circ$ to $360^\circ$ to detect sufficient multi-paths. \rev{As a directional beam of transmitter points to the Rx, the antenna gain of the reflected paths from the ceiling and the floor are 16 dB lower than that of the LoS path. Therefore, the considered reflected paths collected in our experiment are mainly from the desk, chairs and walls, whose elevation angles are sufficiently confined within [-$20^\circ$, $20^\circ$].}

\subsection{Measured Power Delay Angular Profile}
\begin{figure*}

\centering
\subfloat[Path loss over the spectrum of Rx6 in measurement set 1.]{
\includegraphics[width=0.45\textwidth]{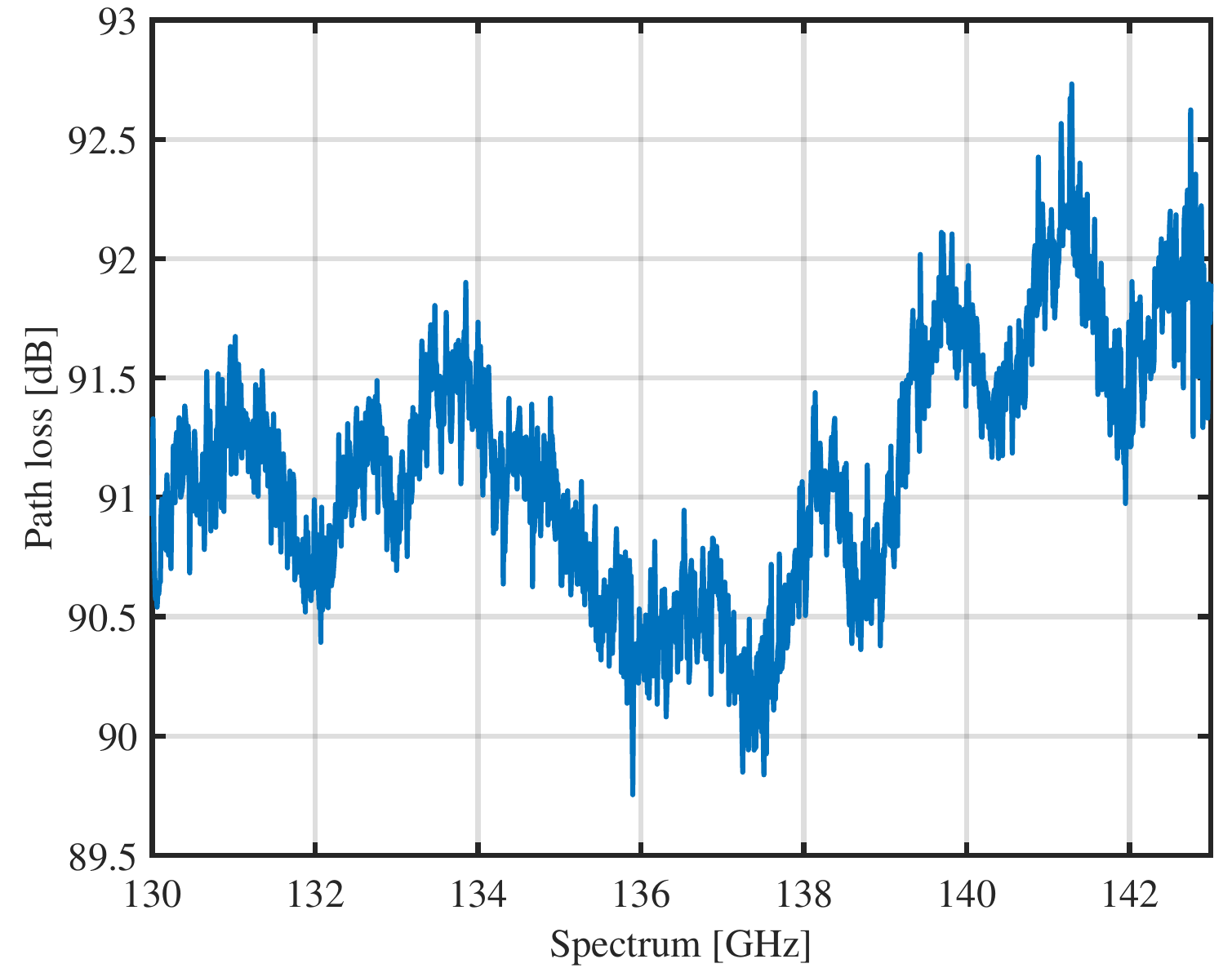}
\label{fig:Opt-Spectrum-6}
%\caption{fig1}
}
\subfloat[Composed 2D PDAP of Rx6.]{
\includegraphics[width=0.45\textwidth]{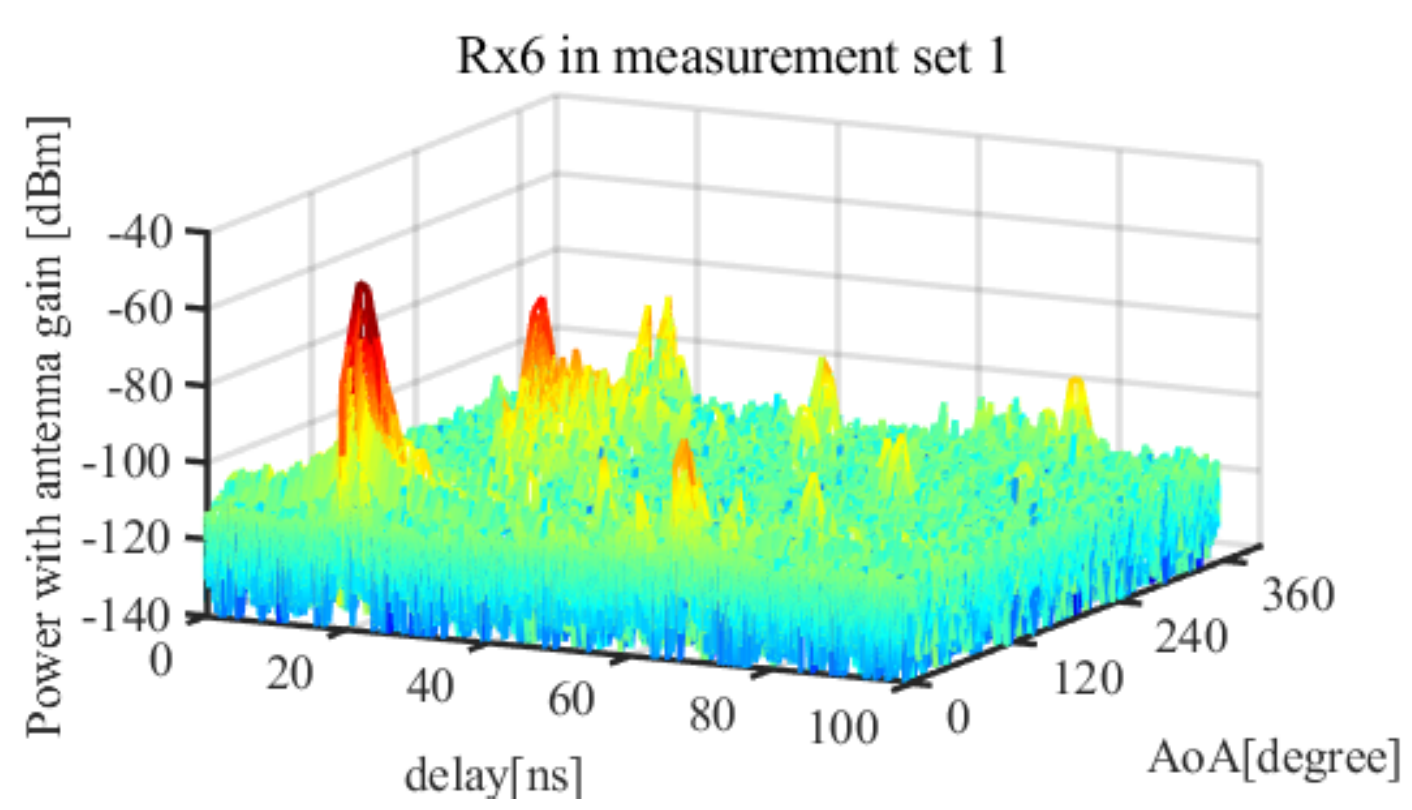}
\label{fig:Com-PDP-sum-6}
}
\caption{Measured path loss and composed power delay angular profile with antenna gain of Rx6 in measurement set 1.}
\label{fig:Rx-6}
\end{figure*}

	\rev{We perform Inverse Discrete Fourier Transform (IDFT) on the transfer function measured from VNA to compute the channel impulse response (CIR) for each azimuth and elevation angle. As a result, we can obtain the three-dimensional (3D) power-delay-angular profile (PDAP), $p_r(i,j,k)$, denoting the received power with time-of-arrival of $i\Delta\tau$, azimuth angle-of-arrival of $j\Delta\theta$, and elevation angle-of-arrival $k\Delta\phi$. Then, we compose all the received power with the same azimuth angle and different elevation angles to obtain the two-dimensional (2D) PDAP of each Tx-Rx pair, calculated as
	\begin{equation}
	\text{PDAP}(i,j)=\sum_k p_r(i,j,k).
	\end{equation}}
\rev{In particular, we do not use the space-alternating generalized expectation-maximization (SAGE) algorithm to estimate the parameters of MPCs for eliminating the effect of antenna pattern and simply regard each point on the PDAP is an MPC. The reason is that the SAGE algorithm in the channel measurement with rotated directional-antenna is highly affected by phase inaccuracy, and requires accurate 3D antenna radiation pattern, which is challenging in practice and time consuming~\cite{60ghzWCX}.}  We take Rx6 as an example and present its path loss over the spectrum and the composed 2D PDAP in Fig.~\ref{fig:Rx-6}\subref{fig:Opt-Spectrum-6} and~\ref{fig:Rx-6}\subref{fig:Com-PDP-sum-6}, respectively. 
\par Since Tx and Rx are well synchronized by the VNA, the delay of the peak in the channel impulse function refers to the absolute time of arrival of an MPC. To verify this, we observe that the delay of the first peak is 19.15~ns, which matches well with the theoretical time-of-arrival (ToA) of the LoS path of 18.4~ns, with the traveling distance of 5.52~m. The minor deviation is caused by a small displacement of the Tx and Rx antennas. \rev{The reason is that the Rx antenna is mounted on a rotator and the rotator is placed on a trolley. When we move the trolley to the marked position, the head of the antenna might not be precisely located at the desired position.}

\subsection{MPCs Clustering and Matching with Ray-Tracing Simulator}

MPCs clustering is to group the MPCs with similar ToA and angle-of-arrival (AoA) into a cluster. Physically, a cluster consists of one reflected path along with some scattered paths surrounding the reflection point. Therefore, the cluster behavior is one of the most significant channel characteristics, based on which many statistical channel models and GSCMs are developed~\cite{gustafson2013mm,ling2017double,li20193d,yang2019cluster}. To cluster MPCs, the unsupervised learning techniques, e.g., K-means and Gaussian Mixture Model (GMM), K-Power-Means (KPM)~\cite{czink2006framework}, Kernel-power-density (KPD)~\cite{he2017kernel} method are utilized. However, those methods except for KPD require manual selection of cluster number and are sensitive to the initial cluster centers. Especially, K-means is limited at handling with the clusters with a ball shape and not flexible for others. Moreover, KPD cannot identify the outliers which are common in our channel measurement results.

\par As the channel measurement only tells the amplitude, angle, and delay information of the MPCs, the physical propagation information of each path is still unknown to us. To solve this problem, we adopt the RT technique to help understand the physical propagation of THz signals in the study of THz channel measurement results. First, RT simulation is conducted in the measurement environment, and then the simulated MPCs from the RT simulator are matched with the MPCs extracted from the measurement. From the literature, one of the most common MPCs matching methods is the threshold-based method~\cite{czink2006novel}. If the multi-path component distance (MCD) between two MPCs, is a measure to quantify the distance between MPCs, is lower than a threshold, the two MPCs are regarded to be matched. Nevertheless, the effectiveness of the MCD matching method is sensitive to the pre-defined threshold.  
\begin{figure*}[htbp]
\centering
\includegraphics[width=0.7\textwidth]{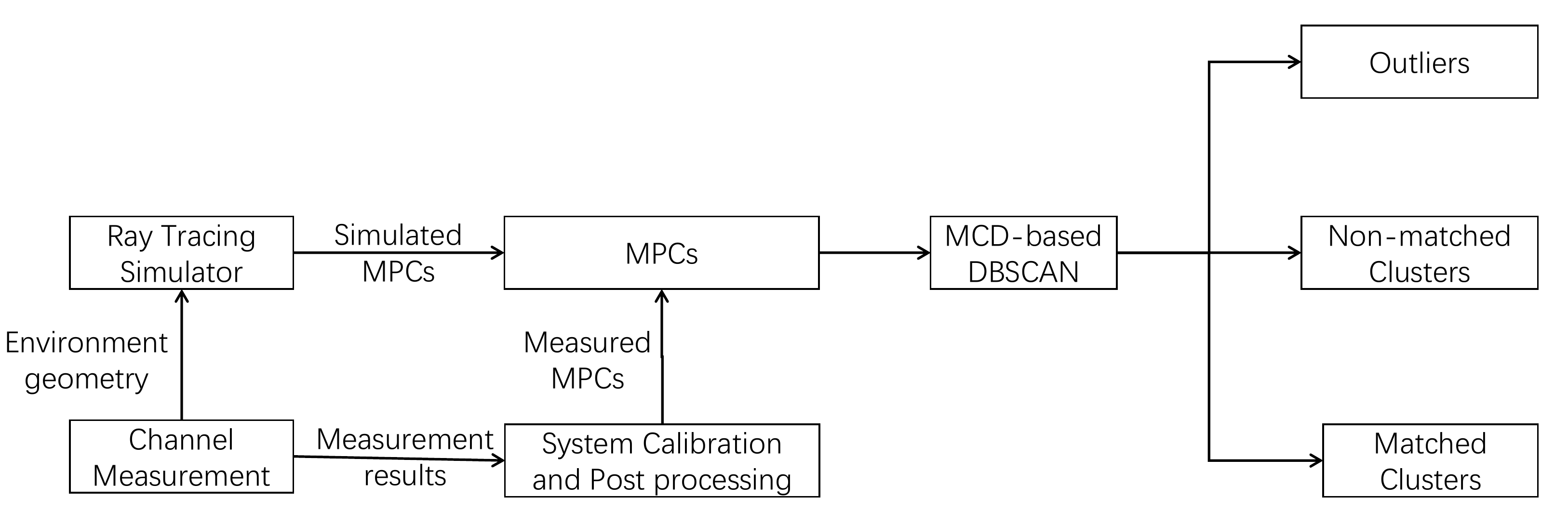}

\caption{Combined MPCs clustering and matching procedure.}
\label{fig:ClusteringMatching}
\end{figure*}

%%%%%%%%%%%%%%%%%%%%%%%%%%%%%%%%%%%%%%%%

%\subsubsection{MPCs clustering and matching procedure} 
Instead, we propose a clustering and matching procedure that combines the MPCs clustering and MPCs matching with RT simulation, with reduced complexity and good clustering performance. The core of clustering and matching in our analysis is MCD-based DBSCAN, which is a density-based clustering algorithm~\cite{ester1996density}, for the following benefits. First, DBSCAN can automatically select the number of clusters, without pre-defining. Second, DBSCAN can find an arbitrarily shaped cluster, which is not restricted to a certain shape like K-means. Third, DBSCAN is robust to outliers, which classifies the MPCs that do not belong to any cluster as noise. By contrast, K-means, GMM, KPM and KPD assign every MPC to a cluster, which causes over-clustering in practice. Fourth, DBSCAN is not sensitive to the initial cluster centers. On the downside, DBSCAN is not good at addressing the dataset where the clusters are close to each other. Fortunately, the THz channel is sparse in both temporal and angular domains, which is suitable to invoke DBSCAN. Furthermore, DBSCAN requires a minimum number of points in a cluster and measures the separation between two points the Euclidean distance. The Euclidean distance is not suitable for measuring the distance in a circular domain (AOA in our case). \rev{Therefore, we replace the Euclidean distance in the original DBSCAN algorithm by MCD, which is given by, 
\begin{equation}
\text{MCD}=\sqrt{||\text{MCD}^2_{\text{AoA}}||+\zeta\text{MCD}^2_{\tau}},
\end{equation}
where $\text{MCD}_\text{AOA}$ and $\text{MCD}_\tau$ are the MCD of AoA and MCD of ToA, respectively.} In addition, $\zeta$ is a suitable delay scaling factor.
In particular, the MCD of AoA is obtained as
\begin{equation}
\begin{split}
\text{MCD}_{\text{AoA}}=\frac{1}{2}|
\begin{pmatrix}
\cos(\theta_1)\\
\sin(\theta_1)
\end{pmatrix}
-
\begin{pmatrix}
\cos(\theta_2)\\
\sin(\theta_2)
\end{pmatrix}|,
\end{split}
\end{equation}
where $\theta_1$ and $\theta_2$ are the AoAs of two MPCs, respectively. MCD of ToA is calculated as
\begin{equation}
\text{MCD}_\tau=\frac{\tau_{std}}{(\Delta\tau_{\text{max})^2}}|\tau_1-\tau_2|,
\end{equation}
where $\tau_1$ and $\tau_2$ are the ToAs of two MPCs, respectively. \rev{Moreover, $\tau_{std}=20$ ns denotes the standard deviation of ToAs of all the MPCs, and $\Delta\tau_{\text{max}}=\text{max}(\tau_l)-\text{min}(\tau_l)=100$ ns.}

\par The flow of the combined MPCs clustering and the matching procedure is depicted in Fig.~\ref{fig:ClusteringMatching}. \rev{In order to filter out the noise in the measured channel as much as possible, we first filter out the noise whose power is lower than a constant threshold of -140 dBm, and then set the minimal number of subpaths in a cluster as 6 in the DBSCAN algorithm. Therefore, if an MPC does not have at least other 5 surrounding MPCs, it does not belong to a cluster and instead, is regarded as noise.} The simulated MPCs from ray tracing and measured MPCs are clustered by the MCD-based DBSCAN algorithm. To be concrete, the output of MCD-based DBSCAN is divided into three parts, including matched clusters, non-matched clusters, and outliers. First, the matched clusters are the clusters consisting of at least one simulated MPC. Second, the non-matched clusters are the clusters consisting of no simulated MPC. The reasons for the appearance of non-matched clusters are twofold. On one hand, the environment geometry fed into the RT simulator is simplified and differs from the real room. On the other hand, the types of traced MPCs depend on the configurations in the RT simulator. In our simulation, since we trace the reflected paths from walls and the non-matched clusters are the paths that reflect from obstacles like chairs and desks. Third, the outliers refer to the MPCs with negligible received power, or the non-matched MPCs from the RT simulator, which are excluded from the channel analysis and modeling. 

\begin{figure*}[htbp]
\centering
\subfloat[\rev{Clustering by the KPM algorithm.}]{
\includegraphics[width=0.45\textwidth]{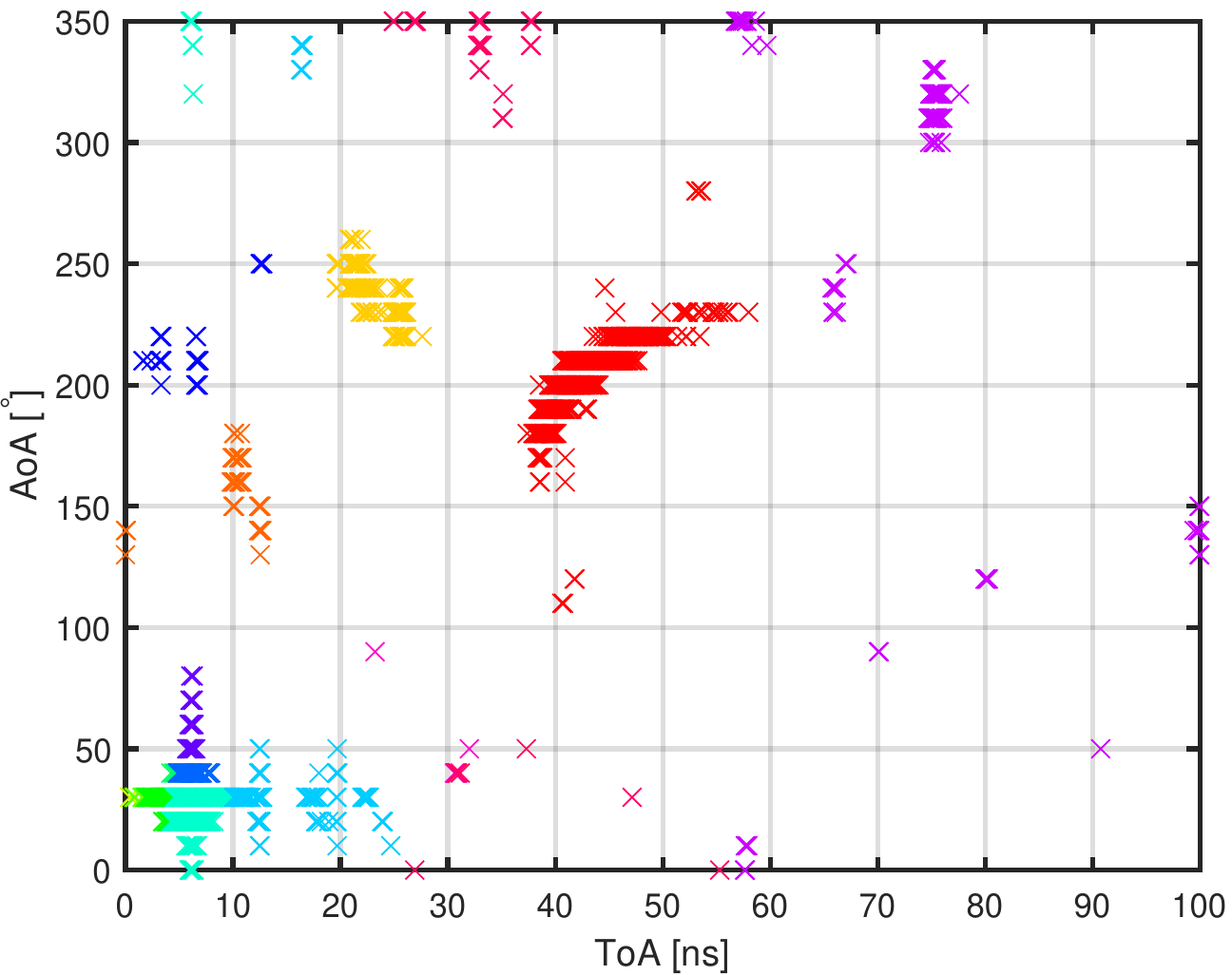}
\label{fig:kpowermeans}
}
\subfloat[\rev{Clustering by the MCD-based DBSCAN procedure. }]{
\includegraphics[width=0.43\textwidth]{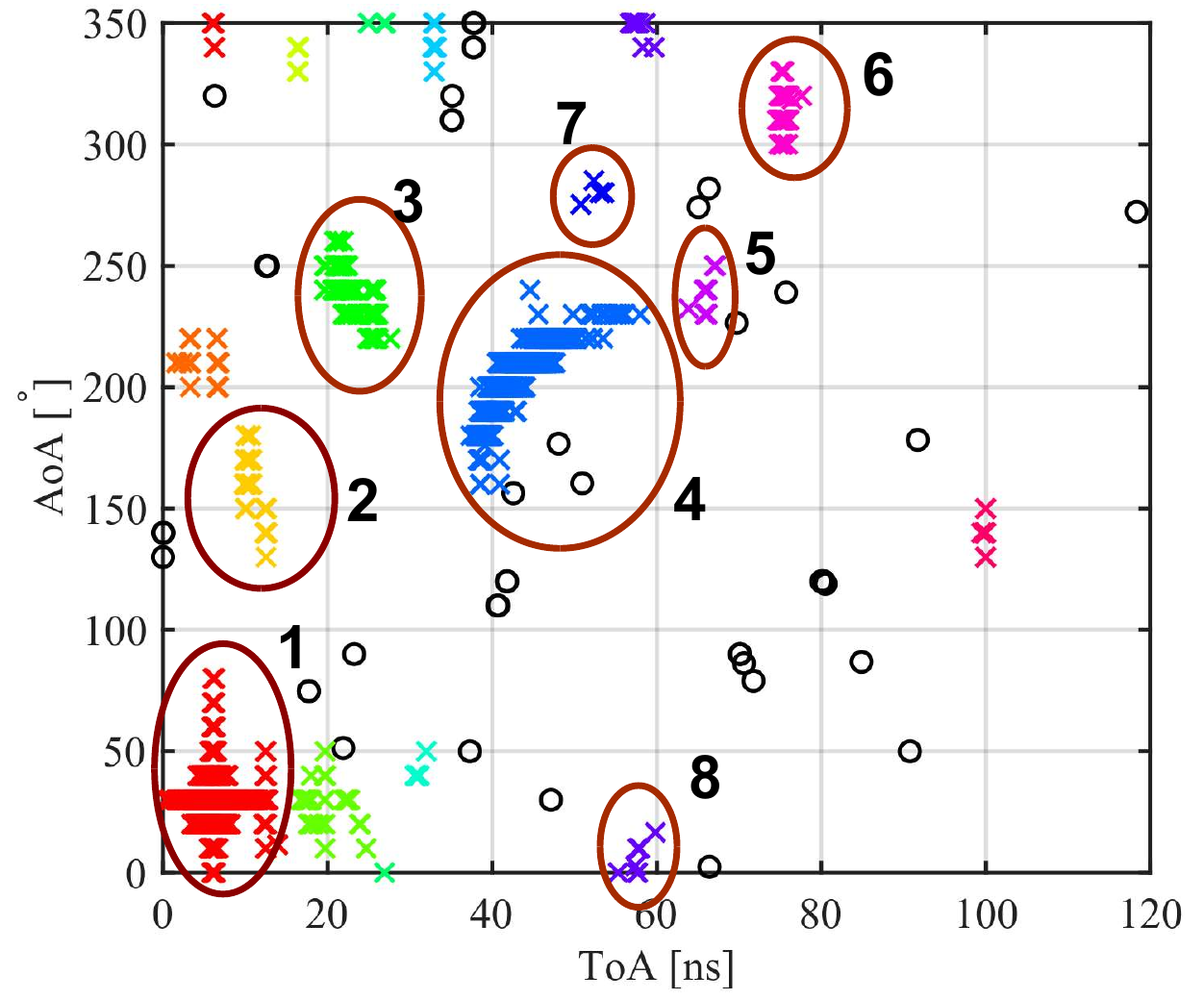}
\label{fig:DBSCAN} 
}
\caption{\rev{Comparison between the K-Power-Means (KPM) algorithm and the proposed MPCs clustering and matching procedure for Rx1. A cross represents an MPC while a circle is noise. Clusters are marked with different colors. Subpaths in a cluster are with an identical color. Some of the matched clusters in (b) are circled with numeric labels.} }
\label{fig:ClusteringResults}
\end{figure*} 

\par \rev{Fig.~\ref{fig:ClusteringResults} illustrates the clustering outcomes of the KPM and the proposed MCD-based DBSCAN algorithms for Rx1 in the measurement set 1. \rev{In Fig.~5, markers with different colors denote different clusters. A cross represents a measured MPCs while a circle represents a noise in Fig.~5(b)}. The number of clusters is set to be the same for a fair comparison.  We observe that KPM groups many MPCs with sparse separation into one cluster in Fig.~\ref{fig:ClusteringResults}\subref{fig:kpowermeans}, due to the drawback that the KPM cannot find outliers. Moreover, KPM splits the LoS cluster into several small clusters, although they are very close. This is due to the fact the number of clusters should meet the pre-determined parameter, and the ball-shape clusters are favored by the KPM algorithm.}
By contrast, Fig.~\ref{fig:ClusteringResults}\subref{fig:DBSCAN} shows that the proposed clustering and matching procedure with MCD-based DBSCAN algorithm outperforms the K-means algorithm. In particular, the matched-clusters are labeled with red circles, and their propagation in the environment is provided by the RT simulator. To be concrete, cluster 1 represents the LoS path with some subpaths caused by the channel measurement platform. Cluster 2 denotes the paths reflected from the desks close to wall 2 in Fig.~\ref{fig:deployment}. Cluster 3 refers to the reflected path from the chairs around the desk in the room center. Cluster 4 and cluster 7 incorporate the first-order reflected paths from wall 1 and wall 2, respectively. Cluster 5 contains the second-order reflected paths traveling through wall 1 and wall 4. Finally, the paths in cluster 6 undergo triple reflections on the wall.
By accurately capturing the clustering effects and matching with the measured MPCs in the indoor environment, the proposed MCD-based DBSCAN algorithm performs well.

%%%%%%%%%%%%%%%%%%%%%%%%%%%%%
\section{THz Channel Characterization and Analysis} 
Based on the processed measurement data, we shed light on the THz indoor channel characteristics, from the perspectives of path loss of the LoS path, MPCs reflection properties, and the temporal and spatial multi-path features in the THz band. The properties and physical parameters revealed in this section are useful as guidelines for THz communication system design. 

\begin{figure}[htbp]
\centering
\includegraphics[width=0.4\textwidth]{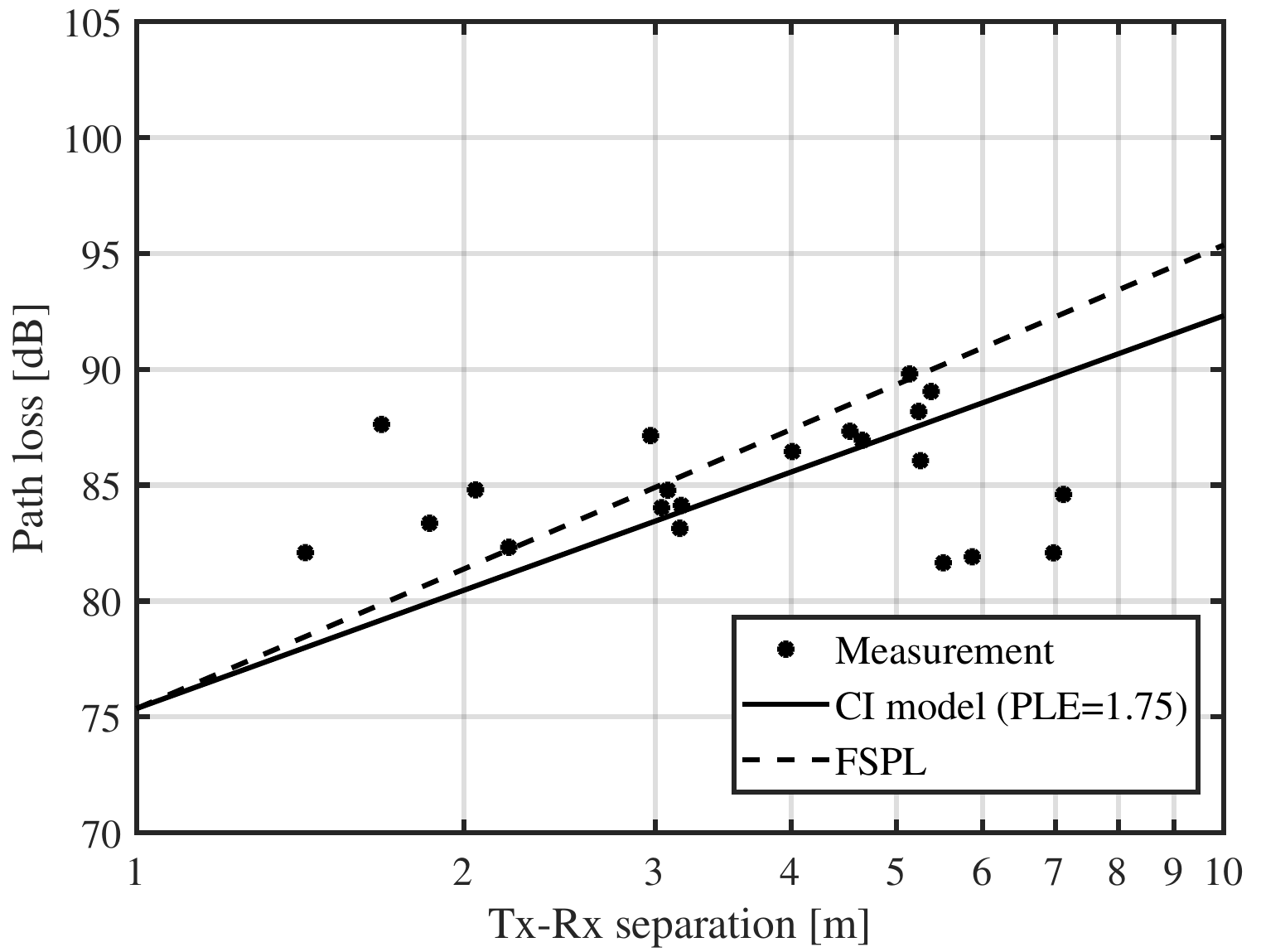}

\caption{Comparison of measured path loss, the fitted CI model and the FSPL results at 140 GHz.}
\label{fig:LoS-PL}
\end{figure}

\subsection{Path Loss Model}
    \rev{The path loss at each Rx in the measurement sets 1 and 2 is calculated by dividing the received power by the transmitted power as depicted in Fig.~\ref{fig:LoS-PL}, where the received power is the sum of the power of multi-paths above the noise floor}. In particular, a \textit{close-in free space reference distance (CI)} path loss model is developed based on the measured path loss, as
    \begin{equation}
    \label{eq:CI}
        \text{PL}^{\text{CI}}\text{[dB]}=10\times\text{PLE}\times\log_{10}(\frac{d}{d_0})+\text{FSPL}(d_0)+X_\sigma,
    \end{equation}
    where PLE is the path loss exponent, $d$ denotes the distance between Tx and Rx, and $d_0$ represents the reference distance which is 1m in this work. $X_\sigma$ is a zero-mean Gaussian random variable with standard deviation $\sigma_{\text{SF}}$ in dB, which represents the fluctuation caused by shadow fading. Moreover, we compute the free-space path loss (FSPL) by invoking the Friis' law, given by,
    \begin{equation}
    \text{FSPL}(d_0,f)=-20\log_{10}(\frac{c}{4\pi fd_0}),
    \label{eq:fspl}
    \end{equation}
    where $c$ denotes the speed of light, $f$ represents the frequency. In addition, PLE in \eqref{eq:CI} is determined by minimizing $\sigma_{\text{SF}}$ via a minimum mean square error (MMSE) approach. The measurement results yield that the PLE value is 1.75 and $\sigma_{\text{D}}$ equals to 3.44~dB for the LoS channel at 140~GHz in a meeting room. 
    
    \begin{comment}
    
    We observe the measured path loss values are well consistent with the FSPL results computed in \eqref{eq:fspl}. This verifies that the calibration process operates properly to suppress system errors and antenna gain effects. 
    Still, an average deviation of 2~dB arises due to the following reasons. First, when the receiver is close to the Tx, for instance, Rx1, Rx2, and Rx4, the measured values of LoS path loss are higher than the results in \eqref{eq:fspl}.
    This is due to possible antenna misalignment at the elevation plane, where the elevation DoA exceeds the scanned range in the measurement campaign.
    Second, constructive effects caused by the reflected path from the desk result in a smaller path loss at Rx8 and Rx9. 
    Third, the largest path loss difference 6.55~dB appears at Rx7, since its LoS path is partially obstructed by a chair. 
    \end{comment}
         \begin{figure}[htbp]
    \centering
    \includegraphics[width=0.4\textwidth]{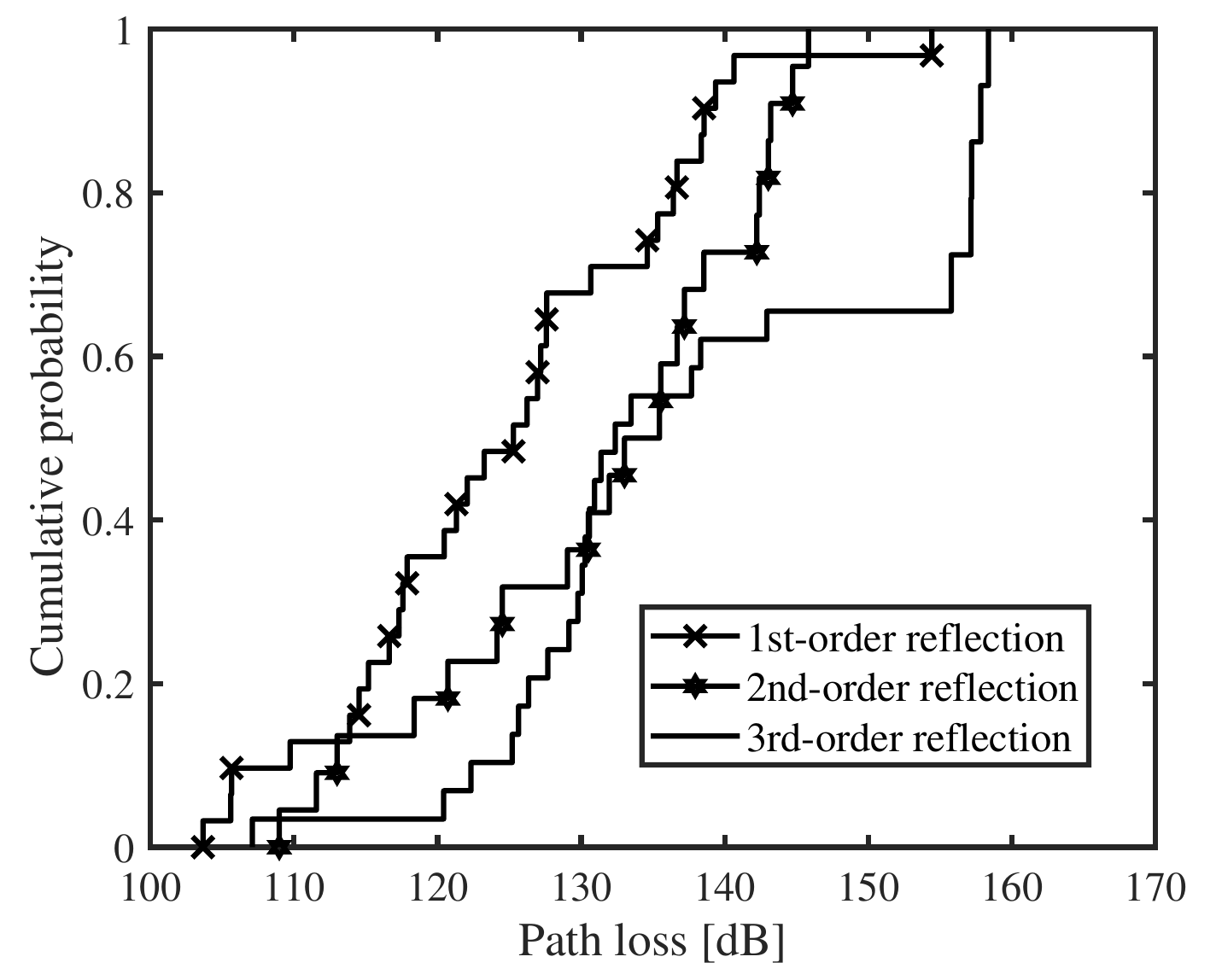}
    
    \caption{CDF of path loss of the reflected paths with different reflection orders at 140 GHz.}
    \label{fig:CDF-ref}
    \end{figure}
\subsection{Reflection Loss}
	The cumulative density functions (CDFs) of the path loss of first-order, second-order, and third-order reflected MPCs in our measurement campaign are computed in Fig.~\ref{fig:CDF-ref}. By including the propagation and reflection attenuation, the average path loss values of the three kinds are 124.2~dB, 129.5~dB, and 138.6~dB, respectively. These results suggest that one reflection leads to additional loss between 6~dB and 9~dB, which agrees with the relative dielectric constant of drywall ($\epsilon_r=6.4$) measured at 140~GHz~\cite{xing2019indoor}.
     Being consistent with intuition, a higher-order reflected path suffers from a higher path loss since it travels through a longer propagation distance and bears excessive attenuation from reflection.
     Indeed, the reflection loss depends on the incident angle, material parameters of objects, and roughness of the reflection surface~\cite{Han2015multiray}.

\subsection{Wall-reflection Paths}
    
     In the indoor environment, we further classify the reflected paths into \textit{wall-reflection} paths and \textit{obstacle-reflection} paths. A wall-reflection path is one that interacts only with the walls inside the meeting room. Otherwise, the reflection belongs to the obstacle-reflection category. The motivation of this classification is that in an extreme case where there are no obstacles like furniture, only wall-reflection paths exist. Note that reflected MPCs from the ceiling and ground are omitted as their elevation angles are out of the scope of our measurement equipment. 
     On the other hand, the existence of furniture and other obstacles in a room can block some wall-reflection paths and produce multiple obstacle-reflection paths.      
     \par To investigate the power distribution among the MPCs, we calculate the power ratio values between the wall-reflection clusters and the obstacle-reflection clusters, $R_w $, which are listed in Table.~\ref{tab:characteristics1} and Table.~\ref{tab:characteristics2}, for the two measurement sets respectively. The power ratio values at all the receivers exceed one, which indicates that the energy carried by wall-reflection MPCs is higher than that carried by obstacle-reflection MPCs, and consequently, the contribution from wall-reflection rays is important to the received signal strength in the THz WLAN environment. The only exception appears at Rx2 in measurement set 2 since there exist some strong NLoS paths that reflect over the desk and chairs near wall 2.
     We notice that in the measurement set 1, Rx1, Rx2, and Rx4 which are on the same side of the desk as Tx receive multiple first-order obstacle-reflection MPCs that come from the chairs around the desk. The ToAs of those MPCs are much shorter than the wall-reflection MPCs, which results in lower free-space propagation loss. Therefore, the values of $R_w$ at these receivers are small. By contrast, Rx3 and Rx6 are more distant from the Tx, and the wall-reflection multi-paths undergo shorter propagation, which leads to very large $R_w$ values. This observation is well aligned with the fact that the obstacle-reflection MPCs are decided by the relative positions of Tx, Rx, and obstacles. In the measurement set 2, $R_w$ values at Rx2, Rx3, Rx4, Rx5, and Rx6 are much lower than others. The reason is that the screen near wall 4 blocks the first-order wall-reflection path and cause another strong \rev{first-order obstacle-reflection} path for Rx2-6.
\begin{table}[htbp]
  \centering
  \caption{Temporal and spatial characteristics for each Rx in measurement set 1.}
    \begin{tabular}{rrrrrrr}
    \toprule
    \multicolumn{1}{l}{Rx} & \multicolumn{1}{l}{$d$ [m]} & \multicolumn{1}{l}{$N$} & \multicolumn{1}{l}{$K$} & \multicolumn{1}{l}{DS [ns]} & \multicolumn{1}{l}{AS [$^\circ$]} & \multicolumn{1}{l}{$R_w$}\\
    \hline
    \hline
    1     & 1.43  & 15    & 28.10  & 6.02  & 29.48  & 2.32  \\
    2     & 3.16  & 14    & 27.88  & 3.63  & 35.36  & 1.89  \\
    3     & 5.26  & 16    & 11.90  & 3.94  & 51.55  & 21.92  \\
    4     & 2.20  & 15    & 48.80  & 6.54  & 27.56  & 1.87  \\
    6     & 5.52  & 8     & 18.03  & 2.35  & 34.76  & 144.15  \\
    7     & 5.14  & 10    & 8.59  & 11.15  & 40.50  & 5.53  \\
    8     & 5.87  & 7     & 56.43  & 2.54  & 20.98  & 134.11  \\
    9     & 6.97  & 7     & 13.67  & 4.39  & 45.56  & 10.20  \\
    10    & 7.12  & 5     & 10.88  & 7.15  & 41.12  & 2.12  \\
    \bottomrule
    \end{tabular}%
  \label{tab:characteristics1}%
\end{table}%

\begin{table}[htbp]
  \centering
  \caption{Temporal and spatial characteristics for each Rx in measurement set 2.}
    \begin{tabular}{rrrrrrr}
    \toprule
    \multicolumn{1}{l}{Rx} & \multicolumn{1}{l}{$d$ [m]} & \multicolumn{1}{l}{$N$} & \multicolumn{1}{l}{$K$} & \multicolumn{1}{l}{DS [ns]} & \multicolumn{1}{l}{AS [$^\circ$]} & \multicolumn{1}{l}{$R_w$}\\
    \hline
    \hline
    1     & 5.28  & 6     & 6.01  & 8.58  & 60.88  & 61.34  \\
    2     & 3.97  & 8     & 36.27  & 4.48  & 29.63  & 0.86  \\
    3     & 3.04  & 13    & 12.91  & 4.60  & 43.23  & 35.59  \\
    4     & 4.63  & 10    & 7.47  & 5.90  & 55.94  & 47.23  \\
    5     & 3.05  & 12    & 30.51  & 6.74  & 30.62  & 14.90  \\
    6     & 1.68  & 10    & 35.14  & 4.90  & 29.16  & 1.14  \\
    7     & 4.65  & 9     & 8.85  & 4.88  & 49.06  & 43.29  \\
    8     & 3.08  & 8     & 31.89  & 4.93  & 29.04  & 67.46  \\
    9     & 1.86  & 11    & 326.94  & 1.79  & 10.78  & 4.14  \\
    10    & 4.01  & 6     & 22.28  & 3.76  & 30.28  & 128.41  \\
    11    & 5.34  & 6     & 18.90  & 3.39  & 33.05  & 68.71  \\
    12    & 3.17  & 8     & 19.56  & 2.85  & 33.49  & 88.71 \\
    \bottomrule
    \end{tabular}%
  \label{tab:characteristics2}%
\end{table}%

\subsection{Temporal and Spatial Features of THz Multi-path Propagation}
\label{sec:temporal-spatial}
    
   The temporal and spatial distributions are important characteristics of multi-path propagation. In Fig.~\ref{fig:DBSCAN}, we clearly observe that the MPCs are resolvable in both temporal and angular domains. Given in Table~\ref{tab:characteristics1} and Table~\ref{tab:characteristics2}, the number of clusters, $N$, is around ten in both measurement sets, varying with the receiver position. Therefore, the indoor channel at 140~GHz presents \textit{sparsity}. Since the cost of the highly directional beam at Tx is that the LoS blockage might cause a severe service interruption of the THz link, it is inappropriate to neglect multi-path effects although the THz channel is sparse. 
According to our measurement results, the multi-path effect is indeed significant when the LoS path is blocked, where considerable energy is contributed by the NLoS MPCs.
    %Therefore, the multi-path effect cannot be neglected in THz WLAN communications, even when a directional beam is radiated and aligned to a UE.
    
    To characterize the significance of the LoS path, we evaluate the \textit{K-factor}, $K$, which is defined as the ratio between the power of the LoS cluster and the power of the remaining NLoS clusters, as computed in Table~\ref{tab:characteristics1} and Table~\ref{tab:characteristics2} for measurement sets 1 and set 2, respectively. A large K-factor suggests that the LoS path dominates in the channel. With a good LoS condition in the meeting room, we compute that the K-factor value ranges from 8.59 to 56.43 in the measurement set 1, and ranges from 6.04 to 326.94 in the measurement set 2, which change greatly over various receiver positions in the room. The highest K-factor appears at those positions with shorter Tx-Rx separation, where the propagation distances of reflected MPCs are substantially longer than that of the LoS. Taking Rx1 as an example, the strongest reflected path is the one that reflects on wall 1. However, its ToA is 6 times of the Los ToA, and the incident angle is almost normal to the plane, which causes very high reflection loss. Thus, its received power is 25~dB lower than that of the LoS path. The second strongest NLoS paths are the first-order obstacle-reflection paths coming from the desk and chairs. Additional 6 dB free-space path loss is introduced since their ToAs are double of the LoS ToA.
    Meanwhile, the angles-of-departure (AoDs) of the two reflected MPCs are $60^\circ$ away from the LoS path, which leads to a noticeable beam misalignment and introduces further attenuation in antenna gains. We observe that the received power of the two obstacle-reflection paths is at least 30 dB lower than that of the LoS path. 
    To sum up, a longer propagation distance, high reflection loss, and beam misalignment of the reflected MPCs are the three critical factors that cause a high K-factor value.

    \rev{The power dispersion of MPCs in the temporal domain is quantified by \textit{root-mean-square (RMS) delay spread (DS)}, $\tau_{\text{rms}}$, which is calculated as, 
    \begin{equation}
    \tau_{\text{rms}}=\sqrt{\frac{\sum^{N_\tau}_{i=0}(i\Delta\tau-\bar\tau)^2A_c(i)\Delta\tau}{\sum^{N_\tau}_{i=0}A_c(i)\Delta\tau}}
    \end{equation}
    with
    \begin{equation}
    \bar\tau=\frac{\sum^{N_\tau}_{i=0}i\Delta\tau A_c(i)\Delta\tau}{\sum^{N_\tau}_{i=0}A_c(i)\Delta\tau}
    \end{equation}
    where $N_\tau$ is the number of sampled ToAs, and $A_c(i)=\sum_j\sum_k p_r(i,j,k)$ denotes the power delay profile.} Similarly, we compute \textit{angular spread (AS)} to characterize the power dispersion of MPCs in the spatial domain. 
    The DS and AS at the measured positions in the measurement set 1 and set 2 are listed in Table~\ref{tab:characteristics1} and Table~\ref{tab:characteristics2}, respectively. On the one hand, the delay and angular spreads differ among the receivers. However, the deviation is insignificant compared to the metric of K-factor. A small DS value suggests that the received power is confined in the temporal domain, while a small AS value indicates the received power comes from a narrow spatial region, which degrades the spatial degree of freedom. 
    
 	Even with very similar geometry properties, the temporal and spatial features could be distinct. For example, we notice that Rx6 and Rx7 are both at a corner of the meeting room and reside symmetrically about the diagonal of the meeting room.
    Besides, the distance difference from Tx1 to the two receivers is only 0.38~m.
    However, Rx6 and Rx7 have substantially different temporal and spatial characteristics, in terms of K-factor, delay spread, and $R_w$. For example, Rx7 has the largest delay spread of $11.152$ ns and the smallest K-factor. By contrast, the smallest DS and the largest $R_w$ are measured at Rx6. The reason is that there exists a very strong first-order wall-reflection MPC (i.e., reflecting on the wall 1 in Fig.~\ref{fig:deployment}) that arrives at Rx6 shortly after the LoS path. 
    \rev{Therefore, Rx6 has a high K-factor, large $R_w$, as well as small DS. By contrast, at Rx7, the wall-reflection MPC (i.e., reflecting on wall 4), which is expected to be strong, is partially obstructed by a metal cabinet. Therefore, all the reflected MPCs have comparable power, which is significantly weaker than the LoS path. As a result, one can observe a low K-factor, large power dispersion in the time domain, and small $R_w$ at Rx7. In a nutshell, delay spread is severely influenced by the critical NLoS MPCs with low path loss and short ToA.}
    
    %\par \textcolor{red}{One should note that the values of $N$, $K$, DS, AS and $R_w$ in Table~\ref{tab:characteristics1} are different than that in our previous conference paper~\cite{yu2020wideband}. First, we cluster the paths with similar ToA and AoA in this paper while didn't consider the clustering effect in~\cite{yu2020wideband}. Therefore, the cluster power is calculated in this paper rather than the power of the strongest path in a local region in~\cite{yu2020wideband}. Second,  in~\cite{yu2020wideband}, the calculations of DS and AS do not account for those paths with the power 30~dB lower than the LoS path. However, in this paper, we filter out the paths with the absolute path loss higher than 140~dB (a lower threshold). That is the reason why DS and AS is much larger in this paper.}
    Furthermore, we use log-normal distribution to fit the THz temporal and spatial features, including $N$, $K$, DS, AS and $R_w$. The log-mean and log-standard-deviation values for those channel parameters in measurement set 1, set 2 and combined set 1\&2 are computed in Table.~\ref{tab:lognormal}. There is no noticeable difference in the mean and standard deviation values for the characteristic parameters between the measurement set 1 and set 2, which suggests that the two measurement sets are statistically consistent.

\begin{table}[htbp]
  \centering
  \caption{Parameters of log-normal distribution for the THz temporal and spatial features.}
    \begin{tabular}{lrrr}
    \toprule
    & \multicolumn{3}{c}{Measurement set} \\
    \cline{2-4}
          & \multicolumn{1}{l}{1} & \multicolumn{1}{l}{2} & \multicolumn{1}{l}{1\&2} \\
    \hline
    \hline
    $\mu_{\ln{N}}$ [ln($\cdot$)] & 2.30  & 2.16  & 2.22 \\
    $\sigma_{\ln{N}}$ [ln($\cdot$)]& 0.40  & 0.26  & 0.33  \\
    \hline
    $\mu_{\ln{K}}$ [ln($\cdot$)]& 3.01  & 3.12  & 3.07  \\
    $\sigma_{\ln{K}}$ [ln($\cdot$)]& 0.63  & 1.00  & 0.86  \\
    \hline
    $\mu_{\ln{\text{DS}}}$ [ln(ns)] & 1.55  & 1.48  & 1.51  \\
    $\sigma_{\ln{\text{DS}}}$ [ln(ns)] & 0.48  & 0.39  & 0.43  \\
    \hline
    $\mu_{\ln{\text{AS}}}$ [ln($^\circ$)]& 3.56  & 3.51  & 3.53  \\
    $\sigma_{\ln{\text{AS}}}$ [ln($^\circ$)]& 0.26  & 0.43  & 0.37  \\
    \hline 
    $\mu_{\ln{Rw}}$ [ln($\cdot$)]& 2.21  & 3.10  & 2.72  \\
    $\sigma_{\ln{Rw}}$ [ln($\cdot$)]& 1.66  & 1.64  & 1.71  \\            
        \hline 
    $\mu_{\ln{\text{CDS}}}$ [ln(ns)]& -1.01  & -1.11  & -1.06  \\
    $\sigma_{\ln{\text{CDS}}}$ [ln(ns)]& 1.18  & 0.95  & 1.07  \\
        \hline 
    $\mu_{\ln{\text{CAS}}}$ [ln($^\circ$)]& 1.77 & 1.79  & 1.78  \\
    $\sigma_{\ln{\text{CAS}}}$ [ln($^\circ$)]& 0.54  & 0.50  & 0.52 \\

    \bottomrule
    \end{tabular}%
  \label{tab:lognormal}%
\end{table}%
\subsection{Intra-cluster Characteristics}
\begin{figure}[t]

\centering
\subfloat[CDS.]{
\includegraphics[width=0.4\textwidth]{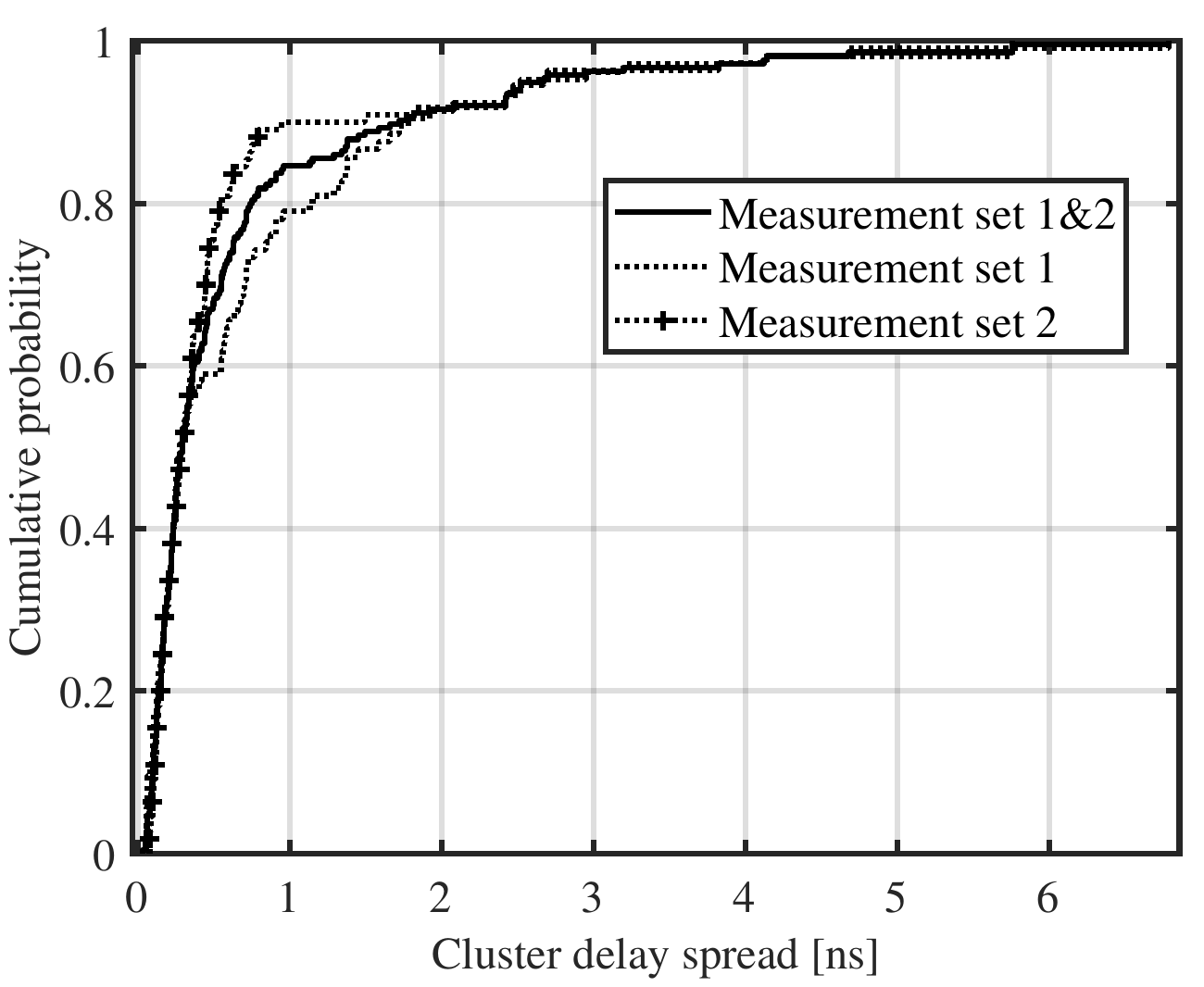}
\label{fig:CDS}
%\caption{fig1}
}
\centering
\subfloat[CAS.]{
\includegraphics[width=0.4\textwidth]{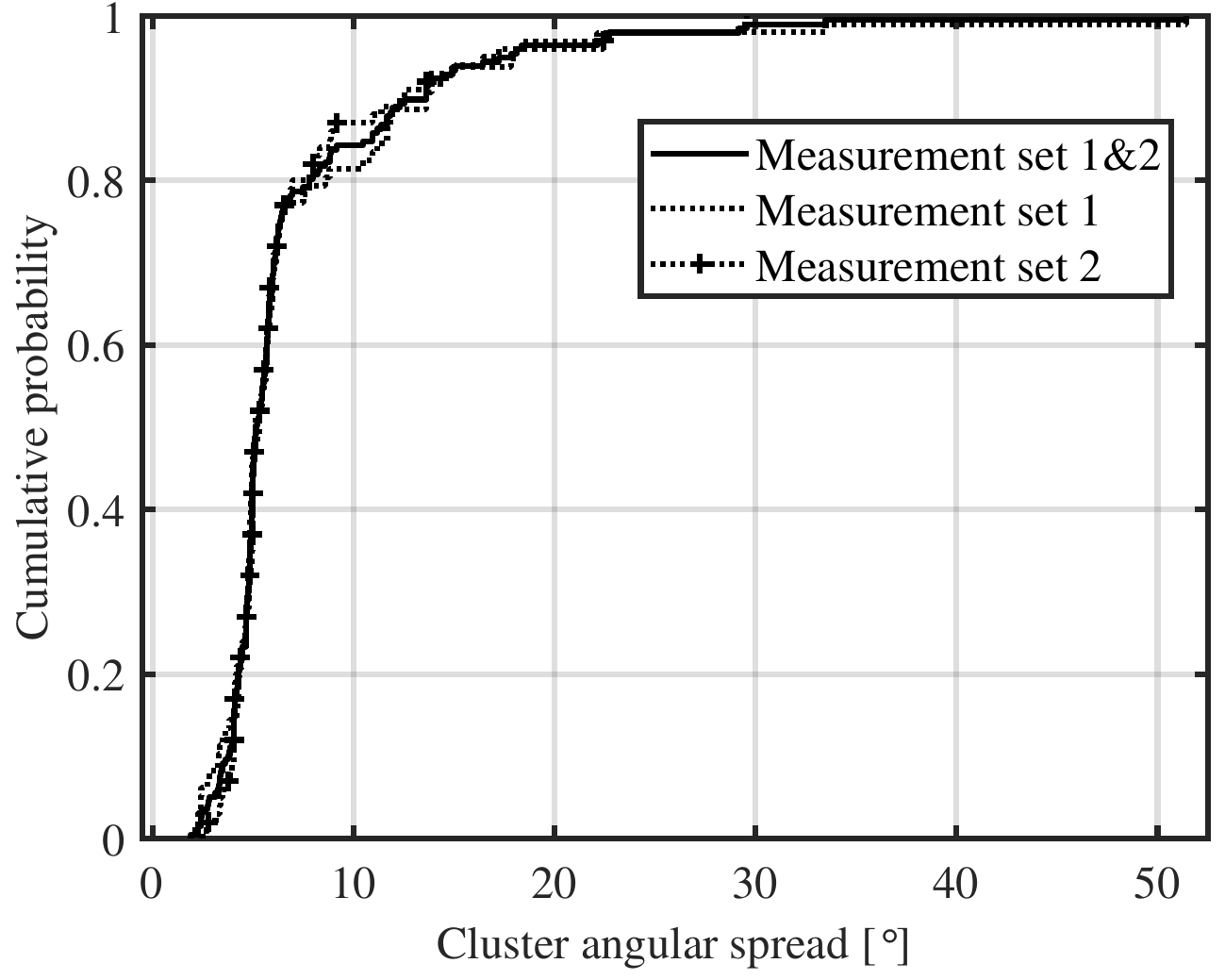}
\label{fig:CAS}
}
\caption{\rev{CDS and CAS distribution of the measured channel at 140 GHz.}}
\label{fig:clustercharacteristics}
\end{figure}

\rev{For intra-cluster characteristics, we analyze cluster delay spread (CDS) and cluster angular spread (CAS) for each cluster. The distributions of CDS and CAS in measurement set 1, set 2 are shown and compared in Fig.~\ref{fig:clustercharacteristics}\subref{fig:CDS} and~\ref{fig:clustercharacteristics}\subref{fig:CAS}. The log-mean and log-standard-deviation values for CDS and CAS in measurement set 1, set 2 and combined set 1\&2 are computed in Table.~\ref{tab:lognormal}. We notice that the mean values of CDS and CAS are both smaller than DS and AS. To be concrete, the mean values of DS and AS are 4.23 ns and $34.12^\circ$, while the mean values of CDS and CAS are 0.35 ns and $5.93^\circ$.}
\begin{table}[htbp]
  \centering
  \caption{Correlation matrix among THz multi-path characteristics.}
    \begin{tabular}{crrrrrr}
    \toprule
          & \multicolumn{1}{c}{$d$} & \multicolumn{1}{c}{$N$} & \multicolumn{1}{c}{$K$} & \multicolumn{1}{c}{DS} & \multicolumn{1}{c}{AS} & \multicolumn{1}{c}{$R_w$} \\
    \hline
    $d$ & 1     & -      &   -    &    -   &  -     & - \\

    $N$ & -0.58 & 1     &     -  & -      &    -   & -\\

    $K$ & -0.39 & 0.12 & 1     &    -   &   -    & -\\

    DS    & 0.12 & 0.04 & -0.35 & 1     &    -   & - \\

    AS    & 0.50 & 0.06 & -0.63 & 0.44 & 1     & -\\

    $R_w$ & 0.29 & -0.50 & -0.16 & -0.42 & -0.05 & 1 \\
    \bottomrule
    \end{tabular}%
  \label{tab:correlmatrix}%
\end{table}%

\subsection{Correlation among THz Multi-path Characteristics}
        To obtain deep understanding and insights of the THz indoor channel properties, we proceed to evaluate the \textit{correlation among the temporal and spatial multi-path characteristics}. In particular, a correlation matrix containing the separation distance among Tx and Rx, the number of clusters, K-factor, delay spread, angular spread, and the power ratio between wall-reflection and obstacle-reflection MPCs, $R_w$, is presented in Table~V. \rev{The element in the correlation matrix is the correlation coefficient between two channel characteristic parameters, which is calculated as,
	\begin{equation}
	\rho_{X,Y}=\frac{\text{Cov}(X,Y)}{\sqrt{\text{Var}[X]\text{Var}[Y]}} 
	\end{equation}
	where $X$ and $Y$ are the two sequences of measured channel characteristics, $\text{Cov}(X,Y)$ denotes the covariance of $X$ and $Y$, and $\text{Var}[\cdot]$ is the variance. The value of the correlation coefficient ranges from -1 to 1. In particular, $\rho_{X,Y}=1$ denotes a positively linear correlation, 0 denotes non-correlation, and $-1$ denotes a negatively linear correlation.}
    \par The main observations are drawn as follows. First, the distance is negatively correlated to the number of clusters and K-factor, which suggests that a long separation distance results in a higher channel sparsity and a lower K-factor. This can be explained that if Tx and Rx separate far, the difference in propagation distance between the LoS and NLoS MPCs decreases. Therefore, for those receiver points far from Tx, the power of LoS is weakened compared with NLoS MPCs.
    
    Moreover, the number of clusters and K-factor demonstrate weak correlation, as their correlation coefficient is equal to 0.12. Furthermore, the number of clusters is neither correlated with delay spread nor to angular spread. This suggests that the number of clusters has no impact on power distribution among the MPCs as well as power dispersion in the temporal and spatial domains.
    Third, the correlation coefficient between delay spread and angular spread is 0.44, which implies that the temporal and spatial spreads are related. This result coincides with the intuition that the power is likely to disperse in the angular domain if it is dispersive in the time domain. This is further aligned with our discussions about Rx6 and Rx7 in Sec.~\ref{sec:temporal-spatial}, where the two receivers with almost the same separation distance have significantly different characteristics. Fourth, $K$ is negatively related to DS and AS, which is consistent with the fact that LoS power is strong with high $K$ and in this case, the power is confined in the temporal and angular domains.
    Fifth, $R_w$ illustrates a positive correlation with the distance. Thus, if a receiver is far away from the transmitter, the wall-reflection power needs to be carefully treated, since it contributes significantly to the received power and dominates over other NLoS MPCs.

    \section{Hybrid Channel Modeling for THz Indoor Propagation}
    The temporal and spatial analysis of the channel measurement results suggest that the critical MPCs, i.e., the LoS path and wall-reflection paths, dominate in the THz indoor channel. Inspired by this, we develop a semi-deterministic or, more precisely, RT-statistical hybrid channel model. The proposed channel model is further validated with the measurement data. Moreover, the performance of the proposed channel model is evaluated and compared with the conventional statistical model and 3GPP channel model~\cite{3gpp2018study}.
    \subsection{Hybrid Channel Model}
    The RT-statistical hybrid channel model generates the channel by combining an RT and a statistical approach. First, we use the RT geometric-optic approach to produce the dominant MPCs, including one LoS path and several wall-reflection paths. Each of these MPCs is the center of a cluster with subpaths. In the RT method, only the dimensions of the room and the positions of Tx and Rx are required. Second, a statistical approach supplements intra-cluster subpaths in the wall-reflection clusters as well as the additional obstacle-reflection clusters. 
    Therefore, the spatial channel impulse response, $h(\tau,\theta,f)$, consists of the deterministic CIR generated by the RT approach, $h_{RT}(\tau,\theta,f)$, and the statistical CIR, $h_{s}(\tau,\theta,f)$, given by
    \begin{equation}
    h(\tau,\theta\rev{,f})=h_{RT}(\tau,\theta\rev{,f})+h_{s}(\tau,\theta\rev{,f}).
    \end{equation}
    where $\tau$ and $\theta$ denote the delay and azimuth AoA. We emphasize that the AoD is not considered since AoD is not scanned in the measurement campaign, as described in Sec.~\ref{sec:deployment}. Also, as elevation AoA is scanned from $-20^\circ$ to $20^\circ$ for each measurement point, which doesn't occupy the whole elevation domain, we compose the CIRs in the elevation domain. Therefore, elevation AoA is eliminated in the model. 
    \subsubsection{RT channel modeling}
    \par The CIR generated by the RT approach, $h_{RT}(\tau,\theta\rev{,f})$, is represented as,
    \begin{equation}
    \label{eq:RT_CIR}
    \begin{split}
    h_{RT}(\tau,\theta\rev{,f})&=A_t(\vec{\phi}_{LoS})\alpha_{LoS}\rev{(f)}\delta(\tau-\tau_{LoS})\delta(\theta-\theta_{LoS})
    \\&+\sum^{L_{RT}}_{l=0}A_t(\vec{\phi}_{l,0})\alpha_{l,0}\rev{(f)}\delta(\tau-\tau_{l,0})\delta(\theta-\theta_{l,0}),
    \end{split}
    \end{equation}  
    where $A_t(\cdot)$ represents the antenna pattern at Tx. \rev{$\alpha_{LoS}\rev{(f)}$, $\tau_{LoS}$, $\theta_{LoS}$ and $\vec{\phi}_{LoS}$ denote the amplitude gain, ToA, azimuth AoA and AoD vector of the LoS path, respectively.} $L_{RT}$ states the number of RT clusters. As we only consider the wall reflection with up to 3 reflection times, the total number of RT clusters is $L_{RT}=20$ (4 first-order wall-reflection paths, 8 second-order wall-reflection paths, and 8 third-order wall-reflection paths). $\alpha_{l,0}\rev{(f)}$, $\tau_{l,0}$, $\theta_{l,0}$ and $\vec{\phi}_{l,0}$ denote the amplitude gain, ToA, azimuth AoA and AoD vector of the central path in the $l^{\mathrm{th}}$ RT cluster, respectively. 
    
    \par In particular, the amplitude gain $\alpha_{l,0}$ is calculated as,
    \begin{equation}
    \alpha_{l,0}\rev{(f)}=\frac{|\Gamma^{(1)}_{l}\rev{(f)}||\Gamma_l^{(2)}\rev{(f)}|\cdots|\Gamma_l^{(R_l)}|}{4\pi f\tau_{l,0}\rev{(f)}}
    \end{equation}
    where $\Gamma^{(n)}_{l }$ is the $n^{\mathrm{th}}$-order reflection coefficient of the $l^{\mathrm{th}}$ RT cluster, $R_l$ is the reflection times of the $l^{\mathrm{th}}$ RT cluster, and $f$ is the frequency. $\Gamma_l\rev{(f)}$ is frequency-dependent and its detailed calculation can be found in~\cite{Han2015multiray}.
    
\begin{table*}[htbp]
  \centering
  \caption{Statistical parameters of the RT-statistical hybrid model.}

    \begin{tabular}{l|l|l|l|l|l}
    \toprule
    \multirow{3}[2]{*}{Variable} & \multirow{3}[2]{*}{Model} & \multicolumn{4}{c}{Prameter}  \\
    \cline{3-6}      &       & \multicolumn{2}{c|}{RT-cluster} & \multicolumn{2}{c}{Non-RT cluster} \\
    \cline{3-6}         &       & Pre-cursor & Post-cursor & Pre-cursor & Post-cursor  \\
    \hline
    \hline
    \multicolumn{1}{l|}{Number of clusters} & \multicolumn{1}{l|}{Given in (13)} & \multicolumn{2}{c|}{-} & \multicolumn{2}{c}{-}      \\
    \hline
    \multirow{2}[2]{*}{Number of subpaths} & \multirow{2}[2]{*}{$\ln(\cdot)\sim \mathcal{N}(\mu,\sigma^2)$} & $\mu=2.09$ & $\mu=4$ & $\mu=1.76$ & \multicolumn{1}{l}{$\mu=2.07$}  \\
          &       & $\sigma=1.20$ & $\sigma=1.59$ & $\sigma=1.33$ & \multicolumn{1}{l}{$\sigma=1.03$}  \\
    \hline
    \multicolumn{1}{l|}{Inter-cluster time of arrival} & \multicolumn{1}{l|}{Given in (14)} & \multicolumn{2}{c|}{-} & \multicolumn{2}{c}{$\lambda_{inter}=13.12$}  \\
    \hline
    \multicolumn{1}{l|}{Intra-cluster time of arrival} & \multicolumn{1}{l|}{Given in (15)} & $\lambda_{intra}=0.0918$ & $\lambda_{intra}=0.0593$ & $\lambda_{intra}=0.102$ & $\lambda_{intra}=0.1417$  \\
    \hline
    \multirow{2}[2]{*}{Inter-cluster angle of arrival} & \multirow{2}[2]{*}{Given in (16)} & \multicolumn{2}{c|}{$\mu_v=-2.6567$} & \multicolumn{2}{c}{$\mu_v=-1.23$}  \\
          &       & \multicolumn{2}{c|}{$\kappa_v=0$} & \multicolumn{2}{c}{$\kappa_v=0$}  \\
    \hline
    \multirow{2}[2]{*}{Intra-cluster angle of arrival} & \multirow{2}[2]{*}{Given in (16)} & $\mu_v=0$ & $\mu_v=0.5$ & $\mu_v=0$ & $\mu_v=0$  \\
          &       & $\kappa_v=33$ & $\kappa_v=3.5$ & $\kappa_v=16$ & $\kappa_v=16$  \\
    \hline
    \multirow{2}[2]{*}{Inter-cluster amplitude} & \multirow{2}[2]{*}{Given in (17)} & \multicolumn{2}{c|}{$a_{inter}=0.34$} & \multicolumn{2}{c}{$a_{inter}=0.36$}  \\
          &       & \multicolumn{2}{c|}{$b_{inter}=-0.65$} & \multicolumn{2}{c}{$b_{inter}=-0.25$}  \\
    \hline
    \multirow{2}[2]{*}{Intra-cluster amplitude} & \multirow{2}[2]{*}{Given in (18)} & $a_{intra}=0.41$ & $a_{intra}=0.42$ & $a_{intra}=0.39$ & $a_{intra}=0.53$  \\
          &       & $b_{intra}=-0.51$ & $b_{intra}=-0.45$ & $b_{intra}=-0.21$ & $b_{intra}=-0.06$  \\
    \bottomrule
    \end{tabular}%
  \label{tab:statistical}%
\end{table*}%

    \subsection{Statistical Modeling}
    In this subsection, we elaborate the statistical part of the RT-statistical hybrid model. In our hybrid channel model, the subpaths of the RT clusters except the central path and all the non-RT clusters are statistically generated. The statistical CIR, $h_s(\tau,\theta,f)$, is calculated as,
    \begin{equation}
    \begin{split}
    \label{eq:statistical_CIR}
    h_{s}(\tau,\theta,f)=&\sum^{L_{RT}}_{l=1}\sum^{P_l}_{p=-Q_l,p\ne0}A_t(\vec{\phi}_{l,p})\alpha_{l,p}\delta(\tau-\tau_{l,p})\delta(\theta-\theta_{l,p}),\\
    &+\sum^{ L_s}_{q=1}\sum^{{S_q}}_{s=-T_q}\alpha_{q,s}\delta(\tau-\tau_{q,s})\delta(\theta-\theta_{q,s}),
    \end{split}
    \end{equation}
    where $P_l$ denotes the number of pre-cursor intra-cluster subpaths in the $l^{\mathrm{th}}$ RT cluster, while $Q_l$ stands for the number of the post-cursor intra-cluster subpaths in the $l^{\mathrm{th}}$ RT cluster. $\alpha_{l,p}$, $\tau_{l,p}$, $\theta_{l,p}$ and $\vec{\phi}_{l,p}$ denote the amplitude gain, ToA, azimuth AoA and AoD vector of the $p^{\mathrm{th}}$ subpath in the $l^{\mathrm{th}}$ RT cluster, respectively. In particular, the subpath with $p=0$ refers to the central path of the RT cluster, which has been characterized by RT. In addition, $L_s$ is the number of non-RT clusters. $T_q$ denotes the number of pre-cursor intra-cluster subpaths in the $q^{\mathrm{th}}$ non-RT cluster, while $S_q$ describes the number of the post-cursor intra-cluster subpaths in the $q^{\mathrm{th}}$ non-RT cluster. $\alpha_{q,s}$, $\tau_{q,s}$, $\theta_{q,s}$ denote the amplitude gain, ToA, azimuth AoA and AoD vector of the $s^{\mathrm{th}}$ subpath in the $q ^{\mathrm{th}}$ non-RT cluster, respectively.
    \par One should note that the antenna pattern of Tx is not involved in the statistical CIR in \eqref{eq:statistical_CIR}, which is different from the RT CIR in \eqref{eq:RT_CIR}. The reason is that all the statistical parameters of the hybrid model are extracted from the channel measurement with a directional antenna at Tx. Therefore, the effect of antenna pattern at Tx is contained in the generated amplitudes.
 	As an overview, the main statistical parameters of the RT-statistical hybrid model are summarized in Table. IV, which are detailed as follows.

\begin{figure}[htbp]
\centering
\includegraphics[width=0.5\textwidth]{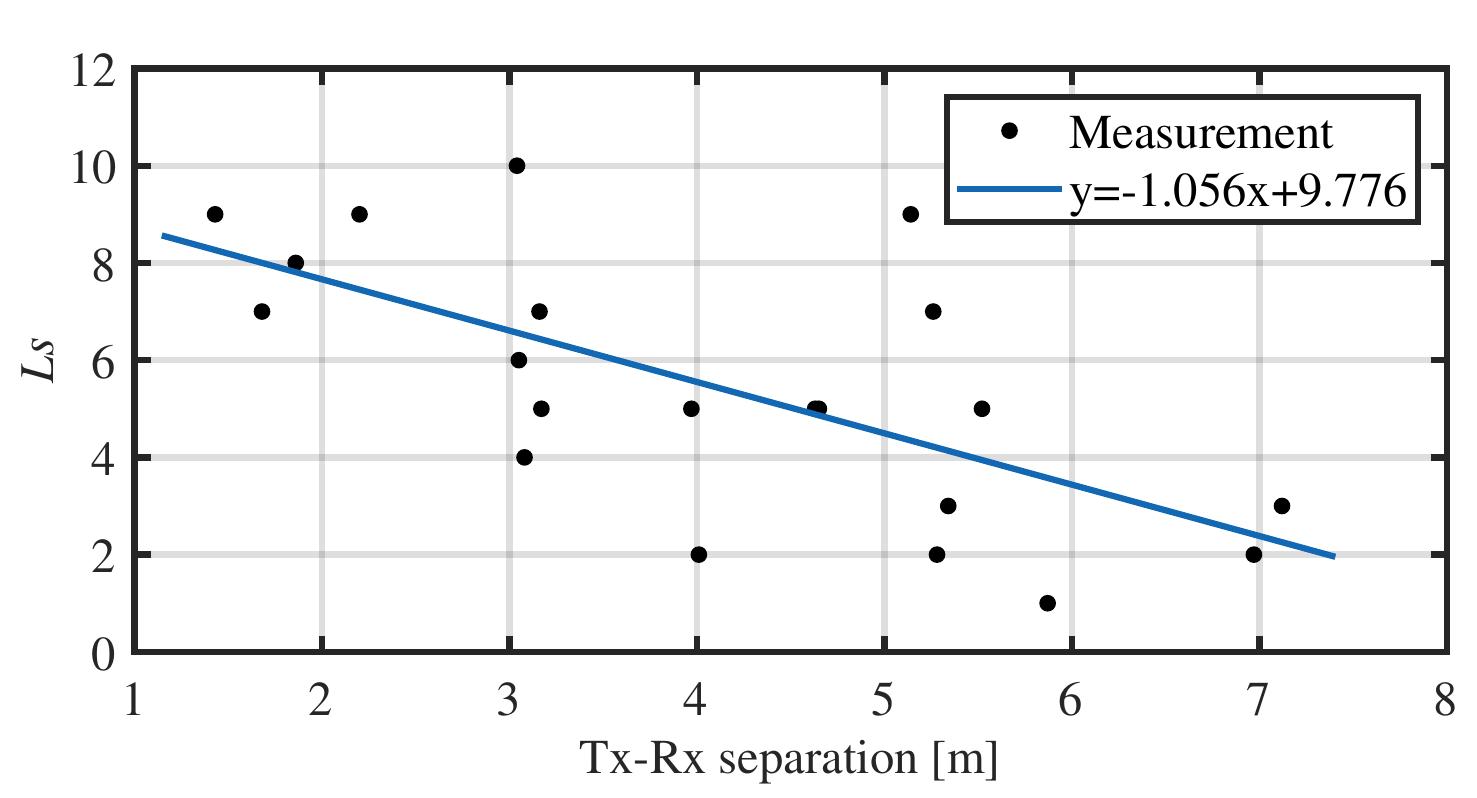}
\caption{Number of non-RT clusters versus the Tx-Rx separation.}
\label{fig:no-distance}
\end{figure}

	\subsubsection{Number of clusters}
	Fig.~\ref{fig:no-distance} shows that the number of non-RT clusters is linearly related to the distance between Tx and Rx. In particular, the channel of far-separated Tx-Rx pair consists of fewer non-RT clusters. Therefore, the number of clusters can be fitted by a linear model as,
    \begin{equation}
    L_s=\lceil-1.056d+9.776\rceil,
    \end{equation}
    where $\lceil \cdot \rceil$ denotes the ceiling function to produce an integer number of clusters.
    \begin{figure}[htbp]
    \centering
    \includegraphics[width=0.5\textwidth]{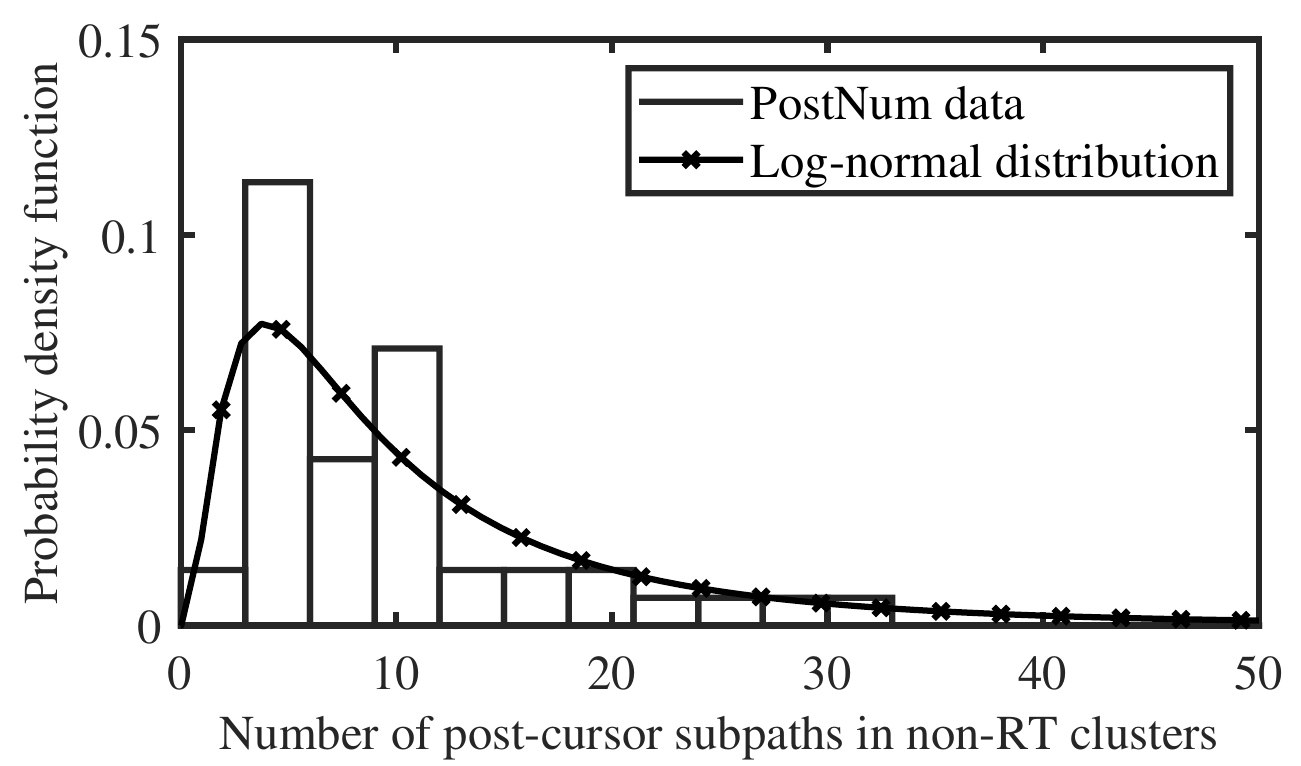}
    
    \caption{Log-normal fitting of the number of post-cursor subpaths in non-RT clusters.}
    \label{fig:NoSubpaths}
    \end{figure}
    \subsubsection{Number of subpaths}
    The numbers of pre-cursor and post-cursor subpaths in both RT and non-RT clusters follow a log-normal distribution. In Fig.~\ref{fig:NoSubpaths}, we present the CDF of the number of post-cursor subpaths in a non-RT cluster and the log-normal fitting, for which the parameters are given in Table IV.
    \subsubsection{ToA}
    The arrivals of inter-cluster MPCs and intra-cluster MPCs are modeled by Poisson processes. Therefore, the ToA between two adjacent inter-cluster MPCs, $\Delta\tau_{inter}$, is independently exponentially distributed, expressed as
    \begin{equation}
    \tau_{l+1,0}=\tau_{l,0}+\Delta\tau_{\text{inter}},\quad\Delta\tau_{\text{inter}}\sim \text{E}(\lambda_{\text{inter}}).
    \end{equation}
    \rev{where $\text{E}(\cdot)$ denotes the exponential distribution and $\lambda_{inter}$ is the inter-cluster rate of arrival.}

    Similarly, the ToA between two adjacent intra-cluster subpaths is independently exponentially distributed as,
    \begin{equation}
    \tau_{l,p+1}=\tau_{l,p}+\Delta\tau_{\text{intra}},\quad\Delta\tau_{\text{intra}}\sim \text{E}(\lambda_{\text{intra}}).
    \end{equation}
    \rev{where $\lambda_{intra}$ is the intra-cluster rate of arrival.}
    \subsubsection{AoA}
    The azimuth AoA of inter-cluster MPCs and the azimuth AoA of the intra-cluster subpaths both follow a Von Mises distribution. The probability density function (PDF) of the Von Mises distribution is given by
    \begin{equation}
f_v(\theta)=\exp(k_v\cos(\theta-\mu_v))/(2\pi I_0(\kappa_v)).
    \end{equation}
     where the parameter $\mu_v$ is a measure of location, $\kappa_v$ is a measure of concentration, and $I_0(\cdot)$ is the modified Bessel function of order 0. \rev{It should be noted that $\mu_v$ and $\kappa_v$ for AoA follow a Von Mises distribution in the unit of rad, as given in Table VI.}

    \subsubsection{Amplitude} The amplitude of the inter-cluster MPCs is calculated as
    \begin{equation}
      \alpha_{q,0}=\alpha_{LoS}a_{inter}|\tau_{l,0}-\tau_{LoS}|^{b_{inter}}.
    \end{equation}
     where $a_{inter}$ and $b_{inter}$ are the coefficient and the exponent for the inter non-RT clusters.

    The amplitudes of the intra-cluster MPCs have the same representation,
    \begin{equation}
    \alpha_{l,p}=\alpha_{LoS}a_{intra}|\tau_{l,p}-\tau_{l,0}|^{b_{intra}}.
    \end{equation}
         where $a_{intra}$ and $b_{intra}$ are the coefficient and exponent of the intra cluster subpaths.

    In addition, the phase of each MPC is assumed to be independently uniformly distributed.
    
    \subsection{Model Validation and Evaluation}
    \label{sec:model_eva}
    We validate the developed RT-statistical hybrid channel model with measured data, and compare it with a conventional statistical model, and the 3GPP TR 38.901 model~\cite{3gpp2018study}.
    In the conventional statistical channel model, the clusters generated by RT are replaced by the statistically-generated ones in which the parameters are adopted from Table~\ref{tab:statistical}. The 3GPP TR 38.901 model is GSCM, which first stochastically generates a virtual map with a certain number of scattering clusters and then constructs the multi-paths based on a simplified RT technique. The GSCM channel model is designed to characterize the angular channel characteristics, and address the problems of smooth time evolution and spatial consistency of the conventional statistical channel model. 
    To generate the GSCM channel, the QuaDRiGa simulator~\cite{jaeckel2014quadriga} is utilized, which is an open-source reference implementation of the baseline 3GPP TR 38.901 model. \rev{Specifically, indoor scenario in 3GPP TR 38.901 model~\cite{3gpp2018study} is chosen for the simulation.} To feed the simulator, the parameters of the 3GPP TR 38.901 model, including path loss exponent, the mean and standard deviation of K-factor, the delay spread, and angular spread as well as the correlation matrix, are taken from our channel measurement for a fair comparison.
    \par \rev{The measured channel data for channel model validation and performance comparison is obtained from a separate channel measurement campaign, where Tx is placed at the position of Rx3 while Rx is placed at the position of Rx7-12 (i.e., 6 measured positions) in Fig.~\ref{fig:deployment}}.

\begin{figure}[t]

\centering
\subfloat[DS.]{
\includegraphics[width=0.4\textwidth]{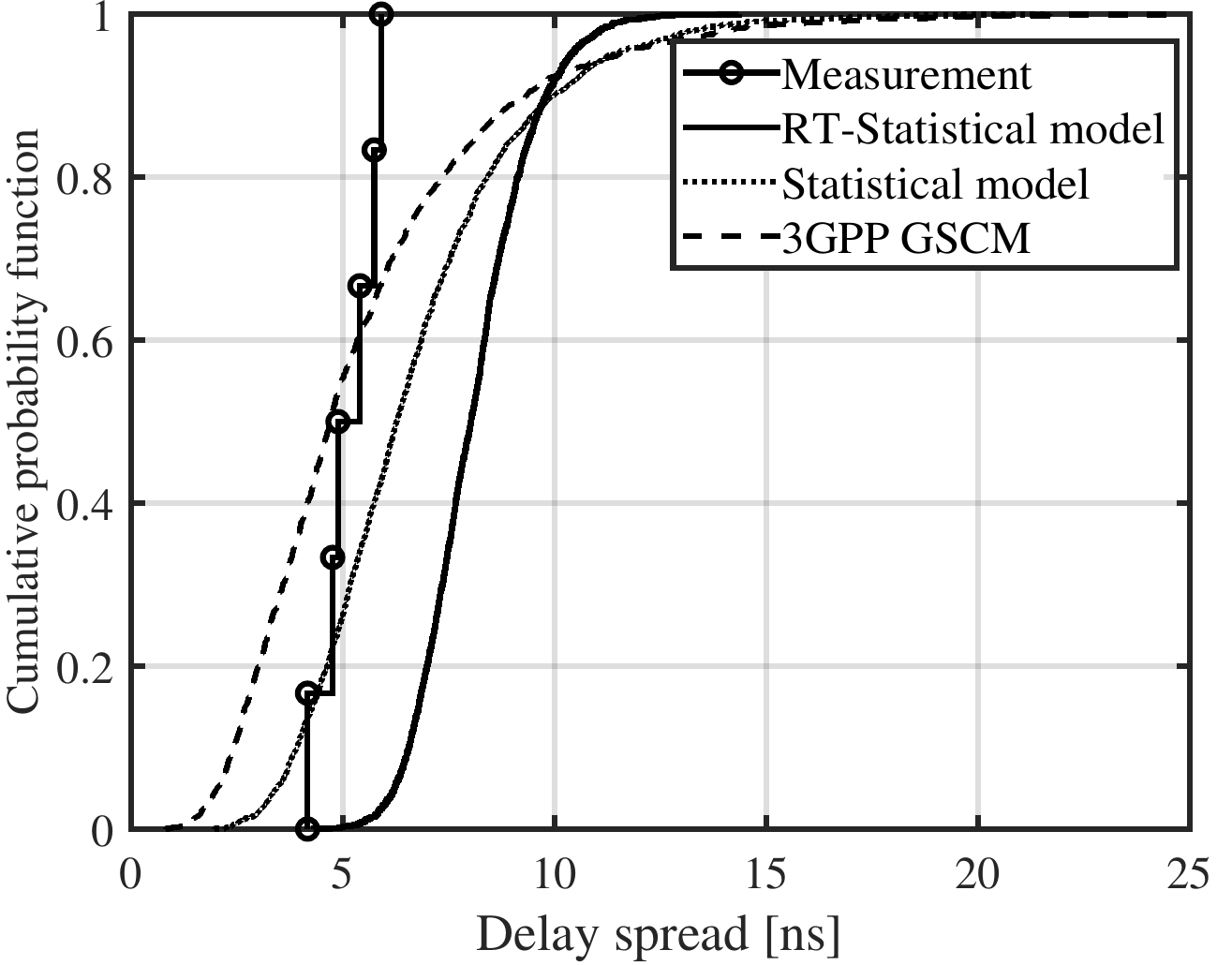}
\label{fig:DS}
%\caption{fig1}
}
\centering
\subfloat[AS.]{
\includegraphics[width=0.4\textwidth]{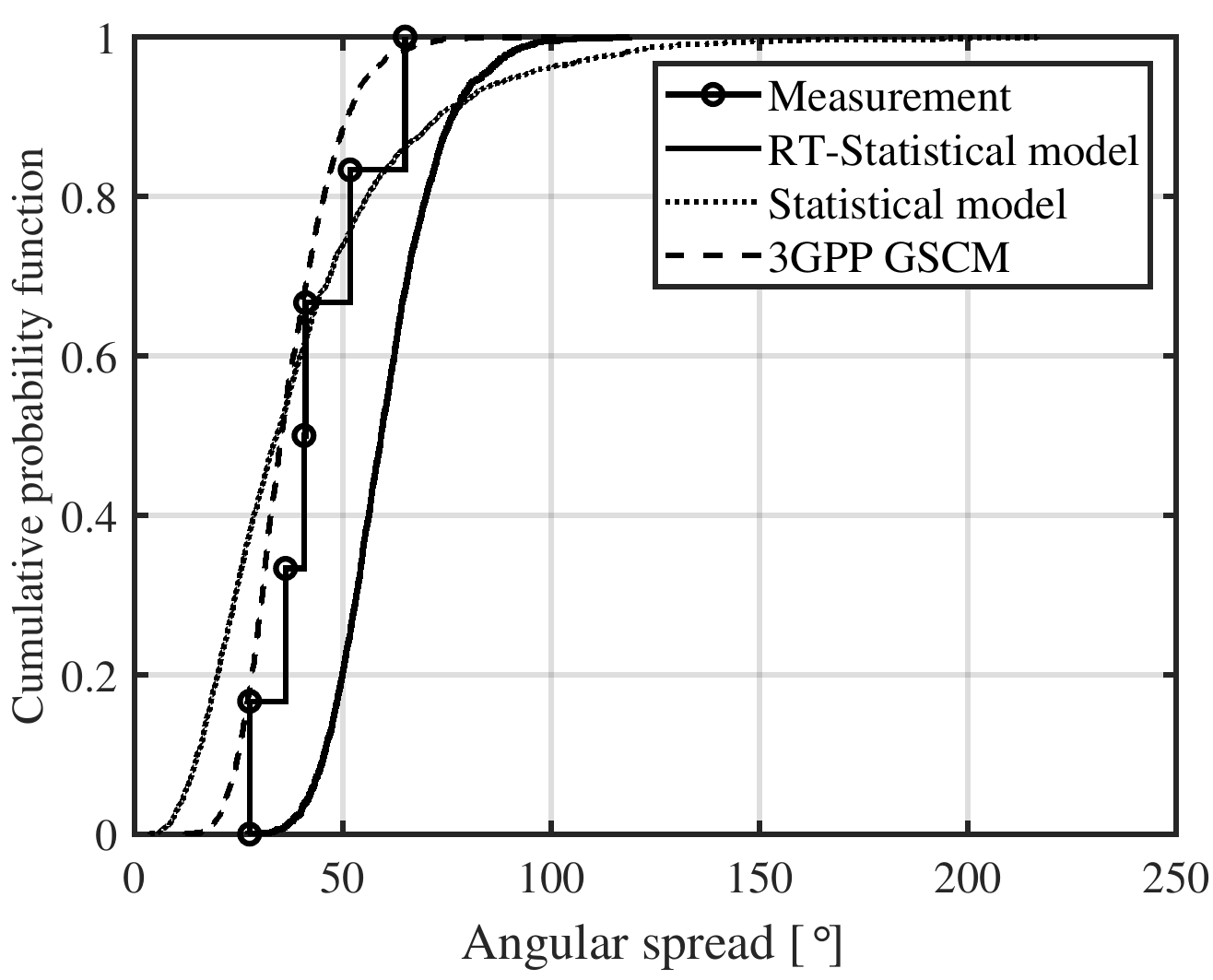}
\label{fig:AS}
}
\caption{DS and AS validation of the proposed RT-statistical channel model and comparison with conventional statistical model and 3GPP GSCM.}
\label{fig:DSAS}
\end{figure}
\vspace{-0.5cm} 
	\subsubsection{Delay Spread and Angular Spread}
	\label{sec:DS}
	The delay spread and angular spread are widely accepted measures of small-scale fading of a channel model. For validation, Fig.~\ref{fig:DSAS}\subref{fig:DS} and~\ref{fig:DSAS}\subref{fig:AS} compare the results of the delay spread and angular spread based on the proposed hybrid channel model, the conventional statistical model, the 3GPP TR 38.901 model, and the channel measurement, respectively. Good agreement is observed, which suggests that 
	the three channel models well characterize the small-scale fading of the THz indoor channel in the temporal domain and angular domains, respectively.

    \subsubsection{Power-delay-angular Profile}
    \label{sec:RMSE}
\begin{figure}[t]

\centering
\subfloat[RMSE of PDAP.]{
\includegraphics[width=0.4\textwidth]{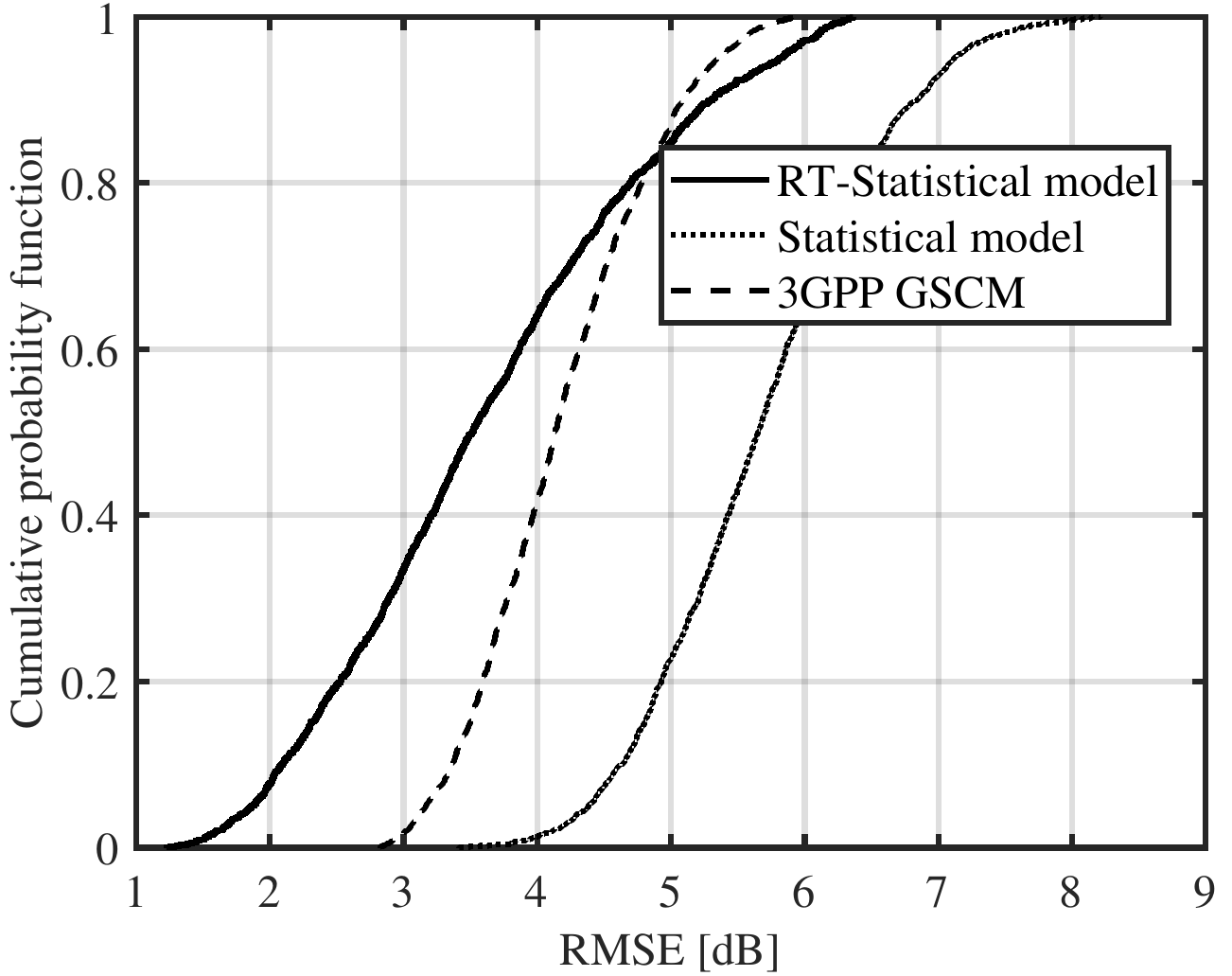}
\label{fig:RMSE}
%\caption{fig1}
}
\centering
\subfloat[SSIM of PDAP.]{
\includegraphics[width=0.4\textwidth]{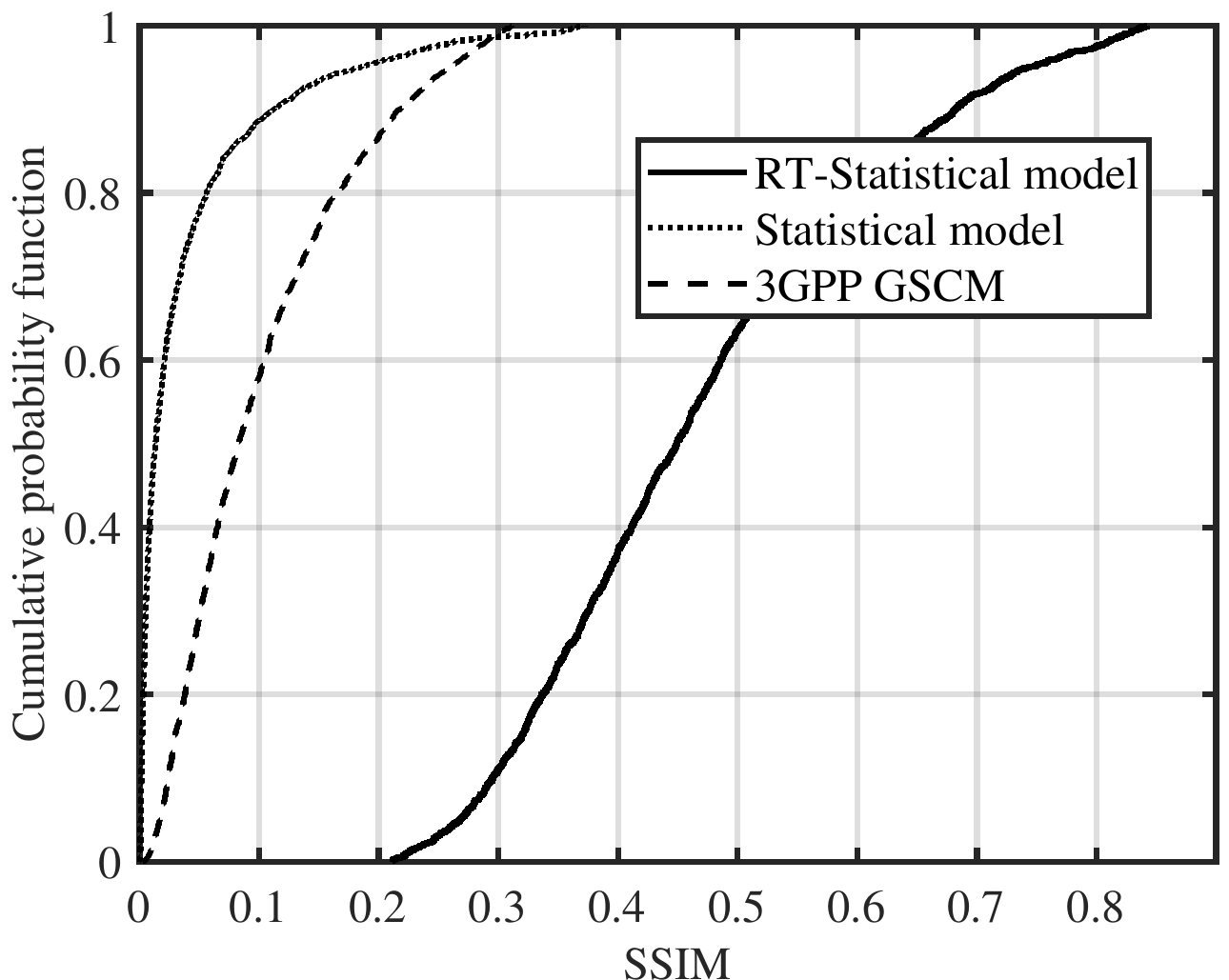}
\label{fig:SSIM}
}
\caption{RMSE and SSIM of the proposed RT-statistical channel model, conventional statistical model and 3GPP GSCM.}
\label{fig:RMSESSIM}
\end{figure}

    The power-delay-angular profile (PDAP) contains the joint temporal-spatial characteristics of the multi-paths of THz channel. We regard that a good channel model should be consistent with the measured PDAP. \rev{The RMSE of the simulated PDAP is calculated by,
	\begin{equation}
	\text{RMSE}=\sqrt{\frac{1}{N_\tau N_\theta}\sum{(\text{PDAP}_m(i,j)-\text{PDAP}_s(i,j))^2}}
	\end{equation} 
	where $\text{PDAP}_m(i,j)$ denotes the measured power with time-of-arrival (ToA) of $i\Delta \tau$ and azimuth angle-of-arrival (AoA) of $j\Delta\theta$. Similarly $\text{PDAP}_s(i,j)$ denotes the simulated power with the time-of-arrival of $i\Delta \tau$ and azimuth angle-of-arrival of $j\Delta\theta$. In addition, $N_\tau$ is the number of sampled ToAs while $N_\theta$ is the number of sampled AoAs. As RMSE calculates the absolute error of the simulated PDAP compared with the measured PDAP, a small RMSE value justifies that the channel model can well characterize the full channel response.} The CDF of the PDAP with respect to RMSE generated by those three channel models is shown in Fig.~\ref{fig:RMSESSIM}\subref{fig:RMSE}. The proposed RT-statistical model has the lowest RMSE, while the conventional statistical model incurs the highest RMSE. \rev{The mean RMSE values of the proposed RT-statistical model, 3GPP GSCM, and conventional statistical model are 3.65 dB, 4.22 dB and 5.73 dB, respectively.}
    \par The RMSE value accounts for the absolute error of all the points between two PDAPs. Therefore, we introduce a Structural Similarity Index Measure (SSIM), which is widely utilized in image compression, image restoration, and pattern recognition as a measure of figure similarity, and has been repeatedly shown to significantly outperform MSE~\cite{wang2010information}. In particular, SSIM is a perception-based model that considers image degradation as perceived changes in structural information variation~\cite{wang2004image}. The range of SSIM is from 0 to 1, which indicates two figures are totally different (i.e., SSIM=0) to exactly the same (i.e., SSIM=1). 
    Fig.~\ref{fig:RMSESSIM}\subref{fig:SSIM} compares the CDF of PDAP SSIM of the different channel models. \rev{The mean SSIM values of the proposed RT-statistical model, conventional statistical model, and 3GPP GSCM are 0.51, 0.05, and 0.14, respectively.} It can be observed that the proposed RT-statistical model has the highest SSIM, which is approximately 10 times higher than the conventional statistical model.  Similar to the metric of RMSE, the 3GPP GSCM is a compromised model. As a result, we state that the proposed RT-statistical hybrid model accurately captures the power distribution in both temporal and angular domains, which performs substantially better than the conventional statistical and the 3GPP GSCM channel models.
 
\vspace{-0.5cm} 
\subsection{Comparison and Discussion}
From the numerical results in Sec.~\ref{sec:model_eva}, we observe that the proposed RT-statistical hybrid channel model, conventional statistical channel model, and 3GPP GSCM can well characterize the delay spread and angular spread of the measured THz indoor channel. However, this does not suffice to validate the accuracy of a channel model.
When it comes to PDAP, the conventional statistical channel model shows the unacceptable performance in terms of RMSE and SSIM. 
The reason is that the conventional statistical channel model only takes the Tx-Rx separation into account, without requiring geometry of the propagation environment as prior information. The ToA and AoA are independently and randomly generated and assigned to each MPC, according to the parameter statistics pre-defined in the statistical model. Therefore, the generated PDAP is not realistic. 

By contrast, the GSCM is semi-deterministic, which considers the positions of Tx and Rx. Therefore, the ToA, AoA, and amplitude of the LoS path in the GSCM are consistent with the measured LoS path. Nevertheless, the scattering environment is still randomly generated according to the pre-defined statistics, before the simplified RT technique is implemented. This leads that the remainder of MPCs is noticeable inconsistent with the measured channel data. As a result, although the 3GPP GBSCM is more deterministic and more accurate than the conventional statistical in terms of PDAP, it still performs worse than our proposed RT-statistical hybrid channel model clearly.

\par The proposed RT-statistical hybrid channel model requires the positions of Tx and Rx, and the dimensions of the room for RT, which introduces additional computational complexity compared with the conventional statistical and the GSCM models. However, the complexity of RT is low since we only trace a small number of LoS and wall-reflection rays with up to triple reflections. 
Moreover, it should be emphasized that the complexity of the statistical part of our hybrid channel model is reduced compared to the conventional statistical model and GSCM counterparts, since the large-scale fading model, i.e., path loss and shadowing, are not necessary. This is due to the fact that the power of LoS and wall-reflection multi-paths dominate in the channel, which is well captured in the RT technique. As a result, both the RMSE and SSIM metrics with respect to PDAP of the proposed model outperform the 
existing models from the literature. 
%We conclude that a little increase in the channel complexity will bring performance improvement in characterizing the temporal and spatial characteristics of the THz indoor channel surprisingly.

\section{Conclusion}
In this paper, we have elaborated extensive channel measurements from 130~GHz to 140~GHz in an indoor meeting room. The directional antenna at Rx is scanned in both azimuth and elevation directions for resolving the multi-paths in the spatial domain. A novel MPC clustering and matching procedure along with ray tracing techniques is utilized to post-process the received multi-path signals. \rev{THz channel characterization and analysis suggest that the LoS and reflected paths from walls dominate in the THz channel of the meeting room, resulting in the measured high K-factor and $R_w$ values.}
%The channel shows a high K factor ranging from 8.59 to 56.43. This indicates that the Los path dominates the THz channel in the meeting room. In addition, the $R_w$ for all the positions exceeds 1, which suggests that the paths reflected by the wall dominate over the other NLoS paths. The reason for the high K-factor and high $R_w$ is that the meeting room consists of very few obstacles, and the obstacles (desk and chairs) are mainly below the height of Rx.
%Due to the high dynamic range of the measurement system, a number of detected clusters are observed in the meeting room, which ranges from 5 to 15. And the number of clusters decreases with the separation between Tx and Rx.
\par \rev{In the meeting room environment, we develop a hybrid cluster-based THz channel model, by combining ray-tracing and statistical methods. The numerical results show that compared with the statistical model and 3GPP GSCM, the hybrid channel model has better agreement with the measured PDAPs for all Tx-Rx positions. This suggests that the deterministic part of the hybrid channel model introduced by ray tracing presents spatial consistency in nature and captures the most significant paths in the THz signal propagation. The statistical part of the hybrid channel model is necessary to complete the multi-path characteristics.}
\bibliographystyle{IEEEtran}
\bibliography{IEEEabrv,CY_bib,bibliography.bib}
\vfill

\end{document}